\documentclass{article}
\usepackage[utf8]{inputenc}
\usepackage{graphicx}
\usepackage{float}
\graphicspath{ {./images/} }
\usepackage{geometry}
\usepackage{changepage}
\usepackage[usenames]{color}

\def\bea{\begin{eqnarray}}
\def\eea{\end{eqnarray}}
\def\be{\begin{equation}}
\def\ee{\end{equation}}

\definecolor{mtcolor}{rgb}{0.13, 0.55, 0.13}

\newcommand{\Sec}[1]{Sec.~\ref{#1}}
\usepackage{amsfonts}
\usepackage{amssymb}
\usepackage{amsmath}
\usepackage{url}
\usepackage{bm}
\usepackage[compress]{cite}
\usepackage{makecell}
\definecolor{rossoCP3}{cmyk}{0,.88,.77,.40}
\definecolor{verdeCP3}{rgb}{0.09765625, 0.57421875, 0.1015625}
\definecolor{bluCP3}{rgb}{0, 0.23, 0.67}
\ifx\pdfoutput\undefined
\usepackage[dvips,bookmarks]{hyperref}    
\else
\usepackage{hyperref}    
\fi
\hypersetup{colorlinks, bookmarksopen, bookmarksnumbered,
citecolor=verdeCP3, linkcolor=bluCP3, pdfstartview=FitH, urlcolor=rossoCP3}

\title{Axion IR}
\author{andrea0292 }
\date{October 2020}

\begin{document}
\begin{center}
{\LARGE\bf Searching for axion-like particle decay in the near-infrared background: an updated analysis\\[1mm] }

\medskip
\bigskip\vspace{0.6cm}

{
{\large\bf Andrea Caputo$^{a,b,c}$,}\
{\large\bf Andrea Vittino$^{d}$,}\ 
{\large\bf Nicolao Fornengo$^{e, f}$,}\
{\large\bf Marco Regis$^{e, f}$,}\ 
{\large\bf Marco Taoso$^{f}$.}\ 
}
\\[7mm]

{\it $^a$ School of Physics and Astronomy, Tel-Aviv University, Tel-Aviv 69978, Israel}\\[3mm]
{\it $^b$ Department of Particle Physics and Astrophysics,
Weizmann Institute of Science, Rehovot 7610001, Israel}\\[3mm]
{\it $^c$ Max-Planck-Institut f\"ur Physik (Werner-Heisenberg-Institut), F\"ohringer Ring 6, 80805 M\"unchen, Germany}\\[3mm]
{\it $^d$Institute for Theoretical Particle Physics and Cosmology (TTK), RWTH Aachen University, 52056 Aachen, Germany}\\[3mm]
{\it $^e$  Dipartimento di Fisica, Universit\`{a} di Torino, via P. Giuria 1, I--10125 Torino, Italy}\\[3mm]
{\it $^f$  Istituto Nazionale di Fisica Nucleare, Sezione di Torino, via P. Giuria 1, I--10125 Torino, Italy}\\[3mm]

\end{center}

\bigskip

\centerline{\large\bf Abstract}
\begin{quote}
\large 
The extragalactic background light 
is comprised of the cumulative radiation from all galaxies across the history of the universe. The angular power 
spectrum of the anisotropies
of such a background at near-infrared (IR) frequencies lacks of a complete understanding and shows a robust excess which cannot be easily explained with known sources.
Dark matter in the form of axion-like particles (ALPs)
with a mass around the electronvolt will decay into two photons with wavelengths in the near-IR band, possibly contributing to the background intensity. We compute the near-IR background angular power spectrum including emissions from galaxies, as well as
the contributions from the intra-halo light and ALP decay,
and compare it to measurements from the Hubble Space Telescope and Spitzer. 
We find that the preferred values for the 
ALP mass and ALP-photon coupling to explain the excess are in tension with 
star cooling data and observations of dwarf spheroidal galaxies. 

\end{quote}

\newpage

\section{Introduction}

The cosmological and astrophysical evidence for the presence of non-baryonic matter, the dark matter, is now overwhelming. Nevertheless, despite a huge theoretical and experimental effort the nature of dark matter is still a mystery. Appealing dark matter candidates are those which can also help to solve other puzzles of the Standard Model (SM) of particle physics. For example, the most promising solution to the strong CP problem involves the introduction of a pseudo-Nambu-Goldstone boson known as the 'QCD axion', which can also play the role of dark matter~\cite{Peccei:1977hh, Peccei:1977ur,Weinberg:1977ma, Wilczek:1977pj, Abbott:1982af, Preskill:1982cy, Dine:1981rt, Dine:1982ah}. A peculiarity of the QCD axion is its coupling to photons, which enables its radiative decay with a lifetime:
\begin{equation}
    \tau_{a} = \frac{64 \pi}{m_a^3 g_{a\gamma\gamma}^2},
    \label{lifetime}
\end{equation}
where $m_a$ is the axion mass and $g_{a\gamma\gamma}$ its effective coupling to photons. Notice that in the case of the QCD axion 
there is a
relation between this coupling and the axion mass, see e.g.~\cite{DiLuzio:2016sbl}. 
More generically, many extensions of the Standard Model, e.g string theory, predict light particles which share the same coupling to photons of the QCD axion, but that might not
be related to
the strong CP problem, and for which $m_a$ and $g_{a\gamma\gamma}$ are independent parameters. These are referred to as axion-like particles (ALPs). The decay of an ALP produces two photons, each with frequency
equal to $\nu = m_a / 4\pi$. 
Depending on the ALP mass various searches of this decay have been proposed, ranging from radio to optical and X-ray frequencies~\cite{Caputo:2018vmy, Caputo:2018ljp,Kephart:1994uy,Rosa:2017ury,Tkachev:1987cd,Sigl:2019pmj,Battye:2019aco,Arza:2019nta,Ghosh:2020hgd,Creque-Sarbinowski:2018ebl,Arias:2012az,Grin:2006aw,Ressell:1991zv,Bershady:1990sw, Bernal:2020lkd}.

In this paper we study the possibility that the ALP decay contributes to the extragalactic near-IR background. In particular, at these wavelengths there is a significant discrepancy between measurements and model predictions for what concerns the angular anisotropies of this background. An even larger discrepancy might be present in the intensity signal, however, currently, its significance is very uncertain mainly due to large uncertainties in the absolute measurements of the foreground contamination by Zodiacal light, see e.g.~\cite{Dwek:2005dj}. We therefore focus on the anisotropy measurements which represent a "cleaner" signal since the Zodiacal light has a smooth spatial distribution, and the angular correlation function of intensity fluctuations is expected to be dominated by extragalactic sources (for sufficiently small angles)\footnote{The possibility to explain the excess in the intensity signal with ALP decays have been studied in~\cite{Kohri:2017ljt,Kohri:2017oqn,Kalashev:2018bra}.}. There seems to be a general consensus that the ``excess'' in the near-IR background anisotropies is of extragalactic origin, even though its nature is still under debate. 
For a recent review on the topic, see~\cite{2018RvMP...90b5006K} and references therein.

\begin{figure}[H]
\centering
\includegraphics[width=0.7\textwidth]{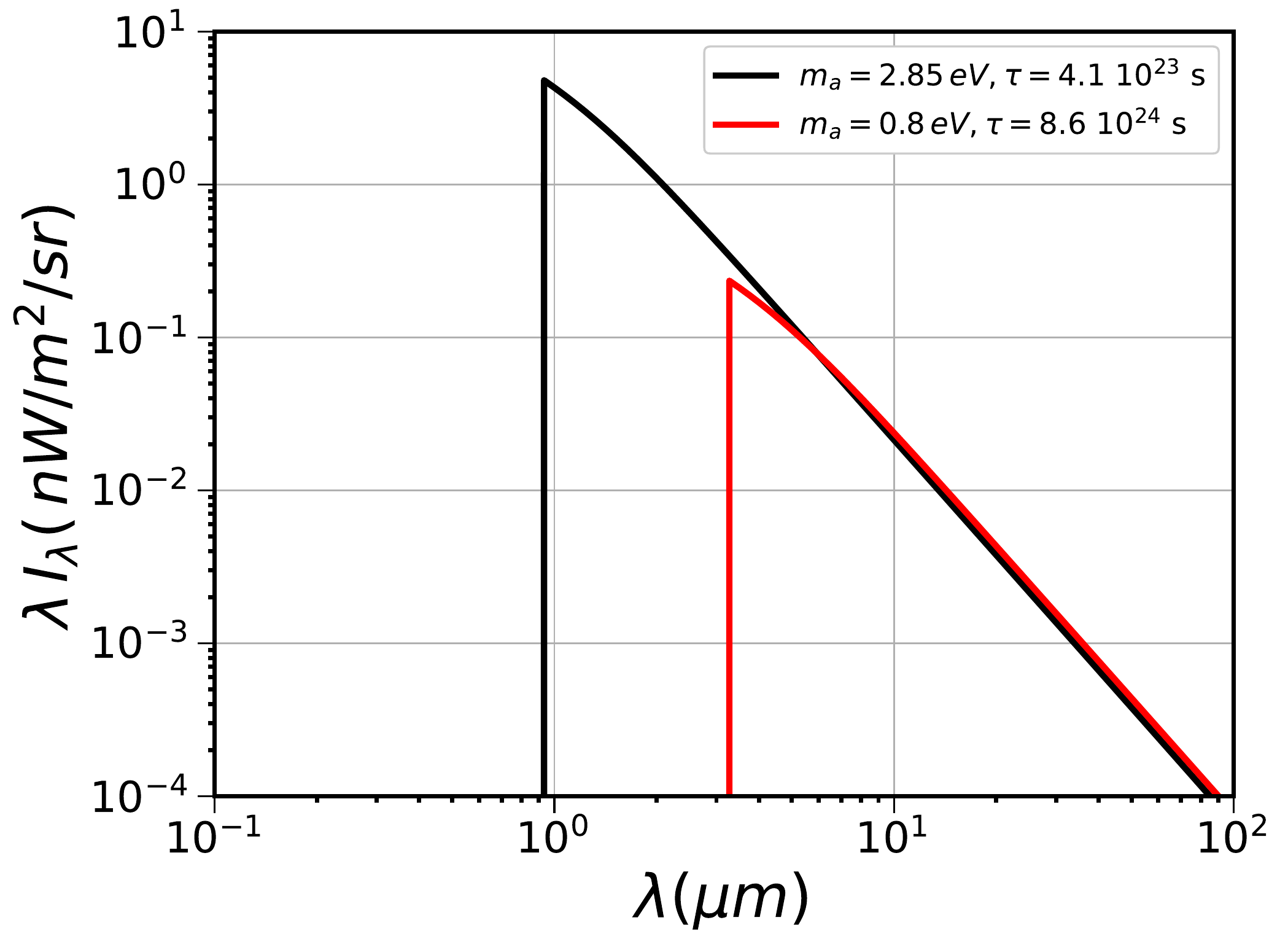}
\caption{The axion-photon decay intensity spectrum as a function of the observed wavelength, for 
$m_a = 2.85$ eV and $\tau = 4.1\cdot 10^{23}$ s (black line), which correspond to $g_{a\gamma\gamma} = 1.2 \cdot 10^{-10} \text{GeV}^{-1}$ and $m_a = 0.8$ eV and $\tau = 8.6 \cdot 10^{24}$ s (red line), for which $g_{a\gamma\gamma} = 1.7 \cdot 10^{-10} \text{GeV}^{-1}$ .}
\label{fig:Axion_Intensity}
\end{figure}

Here we consider as a possible explanation the cosmological IR-emission associated to the ALP decay. The resulting spectrum is composed by a collection of lines at different energies (each line corresponds to a given redshift) which can then contribute to the anisotropy power spectrum of the background and explain the observed excess clustering. 
This possibility was first put under scrutiny in~\cite{Gong:2015hke}, where the authors considered in particular the possibility of having a broad spectrum of ALPs masses. This scenario 
can be realized in string theory constructions, although the typical masses of ALPs predicted are much smaller than those relevant to the near IR emission.A broad spectrum for the dark matter mass can improve the fit to different optical bands, since it adds degrees of freedom with respect to the models with a single ALP mass.  On the other hand, in the present work, we consider the more conventional scenario in which the SM is minimally extended with one ALP of mass $m_a$, and we compare our results with the most recent constraints from other observables, in particular those coming from dwarf spheroidal observations with the optical MUSE-Faint survey~\cite{Regis:2020fhw}. Moreover, we complete the computation done in~\cite{Gong:2015hke} by including the cross-correlation terms among distinct sources of anisotropies, and enlarge the description by considering the contribution coming from the Intra-Halo Light.

The structure of this paper is as follows. 
\Sec{sec:Models} describes the formalism used to compute the cross-correlation signal.
The experimental measurements considered in our statistical analysis are discussed in~\Sec{sec:expdata}. Our results are presented in~\Sec{sec:results}. Finally we conclude in~\Sec{sec:conclusions}.
Technical details are reported in Appendices~\ref{app:High_z} and~\ref{app:fit}.

\section{Models}
\label{sec:Models}
The monopole of the intensity associated to near-IR emission $I_{\rm IR}(E)$  can be written as
\begin{equation}
    I_{\rm IR}(E) = \int_0^{\infty}d\chi W_{\rm IR}(E,z),
\end{equation}
where $\chi(z)$ is the comoving distance (for which $d\chi = c\, dz/H(z)$ in a flat Universe, with $H(z)$ being the Hubble rate), $E$ is the observed energy, while $W_{\rm IR}(E,z)$ is the weight function for the considered source. The weight function quantifies the fraction of intensity emitted at a given energy in a certain redshift slice. Here we consider different IR-sources, namely: low-z galaxies, high-z galaxies, Intra-Halo Light (IHL) and dark matter in the form of ALPs. The intensity of the dark matter decay contribution is shown in Fig.~\ref{fig:Axion_Intensity} for two different values of the axion mass and lifetime. 

The weight function is one of the two ingredients needed to write the angular power spectrum (APS) of the two-point correlation at a given energy:
\begin{equation}
    C_{\ell}^{(i,j)}(E) = \int dz \frac{c}{H(z)}\frac{W_{{\rm IR},i}(E,z)W_{{\rm IR},j}(E,z)}{\chi(z)^2}P_{i,j}(k=\ell/\chi(z),z),
    \label{eq:Cl}
\end{equation}
where $i=j$ describes the auto-correlation case, while $i\ne j$ is associated to the cross-correlation case\footnote{Note that all the considered sources are tracers of the matter distribution on large scales and thus are expected to show some degree of correlation, provided their weight functions overlap.} (i.e., describing the correlation between two different source fields); $P_{i,j}(k,z)$ is the other key ingredient of the APS, giving the three-dimensional power spectrum between the two populations $i$ and $j$. 
We adopt the Limber approximation\cite{1953ApJ...117..134L}, which is valid for $\ell \gg 1$ (i.e., in the multipole range relevant for this work), so that the 3D power spectrum is a function of redshift and modulus of the physical wavenumber $k$, the latter given by $k=\ell/\chi(z)$.

The total APS at a given energy/wavelength is the sum of all auto- and cross-terms:
$C_{\ell}(E)=\sum_{i,j}C_{\ell}^{(i,j)}(E)$.
On top of the APS from physical sources mentioned above and that we model as described in the following two sub-sections, there are other two important contributions that are easier to fit to data, rather than attempting to accurately model. At very small scales, the shot-noise $C_\ell^{\rm shot} = A_{\rm shot}$ dominates the power spectrum, and is given by an amplitude factor $A_{\rm shot}$, which is constant in multipole (for a given observed wavelength), and that we fit to data. 
At very large scales, the APS is, on the other hand, dominated by Galactic foregrounds (mainly diffuse Galactic light), which can be modeled as a power-law~\cite{Mitchell_Wynne_2015}, $C_\ell^{f} = A_f \ell^{-3}$.
The amplitude factor $A_f$ is again left as a free parameter at each wavelength and fitted to the data.

\subsection{Window functions}
For ALP dark matter we consider the emission as given by the ALP decay, described by the window function
\begin{equation}
    W_a(E_{\rm obs},z) = \frac{2\,\Omega_{\rm DM}\,\rho_c}{4\pi\,m_a\, \tau_a}\,E_{\rm obs}\,\delta\left[ (1+z)-\frac{m_a}{2 E_{\rm obs}}\right] \label{eq:W_DM}
\end{equation}
where $m_a$ is the ALP mass and $\tau_a$ is its lifetime. 
$\Omega_{\rm DM}$ and $\rho_c$ are respectively the cosmological dark matter density parameter and the present critical energy density. Notice that the ALP lifetime $\tau$ can be related, for a given mass, to the axion photon coupling via Eq.~\ref{lifetime}.

The photon energy at emission $E_{\rm em}$, being the process a decay into two photons, will simply be $E_{\rm em} = m_a/2$, and so the Dirac-$\delta$ just implies $1+z_{\rm decay} = E_{\rm em}/E_{\rm obs}$. We consider a signal integrated over redshift (see Eq.~\ref{eq:Cl}), 
leading therefore to a continuum spectrum, see Fig.~\ref{fig:Axion_Intensity}
where basically only one redshift contributes for a given wavelength.
Notice that in Eq.~\ref{eq:W_DM} we neglect any broadening of the emission line since the dispersion velocity of the dark matter (typically $< 10^{-3}\,c$ for the massive halos providing the dominant contribution to the signal) is much smaller than the size of the wavelength bin considered in the analysis, which is $> 10$\%, see \cite{Windhorst:2010ib}.

Concerning the contribution from galaxies we consider two different sources: low-z residual galaxies and high-z faint galaxies. For the galaxies at low redshift, we follow Ref.~\cite{Helgason_2012}, where the authors model the extragalactic near-IR emission
arising from known galaxy populations using 233 measured UV, optical and near-IR luminosity functions (LFs) from a variety of surveys. The level of near-IR fluctuation is therefore predicted directly from data,  fitting the parameters of a Schechter-type LF 
\begin{eqnarray}
\phi(M)dM = 0.4 \ln(10) \phi^*\Big(10^{0.4(M^*-M)}\Big)^{\alpha + 1}\times \exp\Big(-10^{0.4(M^*-M)}\Big) dM,
\end{eqnarray}
where $\phi^*$ is the normalization of the LF, $\alpha$ the faint-end slope and $M^*$ the absolute magnitude. To directly compare the flux measurements at different frequencies the authors of Ref.~\cite{Helgason_2012} then adopted the AB magnitude system, which relates the apparent magnitude $m_{AB}$ with the specific flux $f_{\nu}$, via $f_{\nu} = 10^{-0.4(m_{AB}-23.9)}\mu {\rm Jy}$, where $\nu$ is the frequency of the observed photon. Therefore, using the data from various fiducial bands Ref.~\cite{Helgason_2012} provides a fit for $\phi^*(z)$, $\alpha(z)$, $M^*(z)$ and thus a template LFs $\Phi(m|z)$ which fit the data well across a wide range of wavelengths and redshifts. This template luminosity function is then used to compute the galaxy number counts in each magnitude bin per unit of solid angle
\begin{equation}
    N(m) = \int \Phi(m|z)\frac{dV}{dz d\Omega} dz,
\end{equation}
where $dV/d\Omega dz$ is the comoving volume element per solid angle.

The \emph{empirically determined} weight function then reads
\begin{equation}
    W_{\rm low-z}(\nu,z) = \frac{H(z)}{c} \int^{\infty}_{m_{\rm lim}} f(m,\nu) \frac{dN(m|z)}{dz}\,dm,
    \label{eq:low_z}
\end{equation}
where $m$ is the magnitude, $f(m,\nu) = \nu f_{\nu}$ and $m_{\rm lim}$ is the limiting magnitude. This latter parameter separates the resolved and removed galaxies from the unresolved remaining sources.
In the present paper we fix the limiting magnitude by fitting the shot-noise in each band using the relation:
\begin{equation}
C_{\rm SN}(\nu)=\int_0^{z_{\rm max}} dz\,\int^{\infty}_{m_{\rm lim}} dm\,f(m,\nu)^2\, \frac{dN(m|z)}{dz}\;,
\end{equation}
where $z_{\rm max} =7 $ is the redshift threshold adopted in the modeling of Ref.~\cite{Helgason_2012}.
This essentially means to assume that low-z galaxies are the main source of shot noise 
~\cite{KASHLINSKY_2005, Kashlinsky_2007}. Once the limiting magnitude is fixed, the contribution from low-z galaxies is entirely specified. We anticipate here that the shot-noise contribution will not be included in our full Monte Carlo analysis. The reason being two-fold. On one hand, the evaluation of Eq.~\ref{eq:low_z} is computationally demanding and leaving the parameters $m_{\rm lim}$ free at each point of the iteration is unpractical. On the other hand -- and more importantly -- at very small angles (large multipoles) the shot noise will always be the main contribution of the total fluctuations by more than two orders of magnitude. It is therefore justified to first fit the shot noise term using the data at very large multipoles, and then use the inferred value to determine $m_{\rm lim}$.

Another important fraction of the extragalactic near-IR background comes from redshifted photons from the ultraviolet (UV) emission in galaxies at $z>6$, during reionization. In this case we adopt the analytical model of Ref.~\cite{Cooray_2012}, which provides the near-IR background coming from  direct emission from the stars, Lyman-$\alpha$ line, free-free, free-bound and two photon processes. We considered both nebular and intergalactic emissions, but we found the latter to be always subdominant and we finally neglected it. We provide details for the various emission sources in Appendix ~\ref{app:High_z}.

In this case the weight function is given by 
\begin{equation}
\label{eq:Whighz}
   W_{\rm high-z}(\nu,z)= \frac{\nu}{4\pi(1+z)} l_{\nu}\langle\tau_* \rangle \psi(z),
\end{equation}
where $l_{\nu}$ is the luminosity mass density, $\langle\tau_* \rangle$ is the mean stellar lifetime and $\psi(z)$ the comoving star formation density, as given by~\cite{Santos_2002}

\begin{equation}
\psi(z) = f_* \frac{\Omega_b}{\Omega_m}\frac{d}{dt}\int^{M_{\rm max}}_{M_{\rm min}}M\frac{dn}{dM}(M,z).
\end{equation}
Here $f_{*}$ is the star formation efficiency, which we fix to be $f_{*}=0.03$~\cite{Schneider:2018xba, Wise:2014vwa},
while $\Omega_m$ and $\Omega_b$ are the cosmological density parameters respectively for the total matter and baryons.
$dn(M,z)/dM$ is the halo mass function~\cite{Sheth:1999mn}, with $M_{\rm min}$
being the threshold mass for a dark matter halo to form a galaxy during reionization and $M_{\rm max}$ an upper limit for the halo masses. 
We computed the mass function within the halo model and cross-checked our results with the public package Colossus~\cite{Diemer_2018}, finding excellent agreement. Throughout this work, when not specified, we assume $M_{\rm min}= 10^{6}M_{\odot}$ and $M_{\rm max}= 10^{18}M_{\odot}$.

In addition, we also consider the contribution from IHL, coming from diffuse intra-halo stars of all galaxies, which may have originated from tidally stripped stars during galaxy mergers and collisions. 
For the modeling of IHL we closely follow Refs.~\cite{cooray2012measurement} and \cite{Mitchell_Wynne_2015}, which add a diffuse extended component in addition to the standard galaxy clustering models. The IHL luminosity-mass relation is 
\begin{equation}
    L_{{\rm IHL},\lambda}(M,z)=f_{\rm IHL}(M)L(M,z=0)(1+z)^{\alpha}f_{\lambda}(\lambda/(1+z)),
\end{equation}
where $M$ is the mass of the halo, $f_{\rm IHL} = A_f \Big(\frac{M}{10^{12}M_{\odot}}\Big)^{\beta}$ is the fraction of total luminosity in form of IHL, $\alpha$ is the power-law index that accounts for redshift evolution, and $f_{\lambda}$ is the spectral energy distribution (SED) of the IHL. The total luminosity as a function of halo-mass at $z=0$, $L(M,z=0)$, has been taken from the measurements of galaxy groups and clusters~\cite{Lin:2004ak}, and extended to lower masses using the same power-law slope. The IHL SED is taken to be the same as the SED of old elliptical galaxies, as done in Ref.~\cite{cooray2012measurement}. For definiteness we set $\alpha=1$ and $\beta=0.1$. Nevertheless we verified, via a $\chi^2$ minimization, that varying $\alpha$ in the allowed range given in Refs.~\cite{cooray2012measurement} and \cite{Mitchell_Wynne_2015} does not affect the results on axion parameters.
We write $A_f=10^{-3}\,A_{\rm IHL}$, and leave the parameter $A_{\rm IHL}$ free in our fit. 

\begin{figure}[H]
\centering
\includegraphics[width=0.7\textwidth]{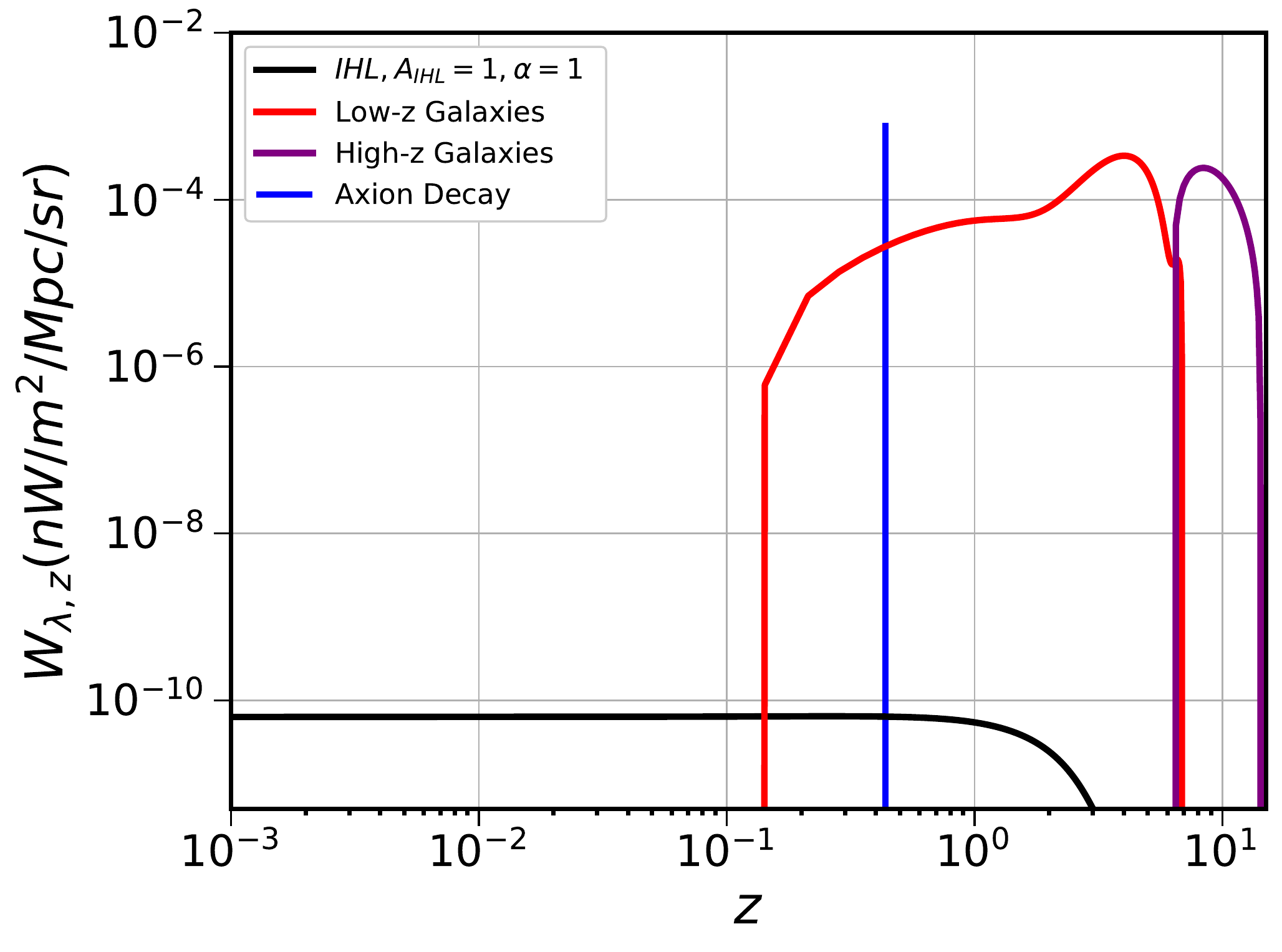}
\caption{Window functions at $\lambda_{\rm obs} = 1.25 \, \mu m$ for the four principal sources considered in this work:  intra-halo light (black), low-z galaxies (red), high-z galaxies (purple) and dark matter decay (blue) for $m_a = 2.85$ eV and $\tau = 4.1 \cdot 10^{23}$ s.}
\label{fig:Window_functions}
\end{figure}

The window function can be written as:
\be
W_I(\lambda,z)=\int_{M_{\rm min}^I}^{M_{\rm max}^I} dM \, \frac{dn}{dM} \,\frac{L_{{\rm IHL},\lambda}(M,z)}{4\pi\,(1+z)}\;.
\ee
We set $M_{\rm min}^I=10^9\,M_\odot\,h^{-1}$ and $M_{\rm max}^I=10^{13}\,M_\odot\,h^{-1}$, as in 
Ref.~\cite{Mitchell_Wynne_2015}. In Fig.~\ref{fig:Window_functions} we show the four window functions: intra-halo light (black), low-z galaxies (red), high-z galaxies (purple) and dark matter decay (red star) for $m_a = 2.85$ eV and $\tau = 4.1 \cdot 10^{23}$ s. Notice that for a given observed frequency and axion mass, the dark matter window function is non zero only at $z_{\rm decay}$, and thus shows a spiky feature.
The IHL contribution comes from low redshift, quickly vanishing for $z>2$. 
The two populations of galaxies have little overlap between $6<z<7$, but essentially they can be treated as separated samples.
Furthermore, the peak in the window function of low-$z$ galaxies is not at very small redshift, which means their average physical distance is significant and their angular size in the sky small. Only the IHL and DM sources can then be seen as extended objects at the angular scales relevant in this work.

\subsection{Three-dimensional power spectra}
In the halo model computation, the 3D power spectrum (PS) is composed by the one-halo ($P^{1h}$) and two-halo ($P^{2h}$) terms with $P=P^{1h}+P^{2h}$.
For a derivation of the $P^{1h}$ and $P^{2h}$ discussed in the equations below, see \cite{Fornengo:2013rga}.
In the case of ALP dark matter, it is given by the dark matter power-spectrum:
\bea
P^{aa}_{1h}(k,z) &=& \int_{M_{\rm min}}^{M_{\rm max}} dM \, \frac{dn}{dM} 
\tilde u^2(k|M)\nonumber \\
P^{aa}_{2h}(k,z) &=& \left[ \int_{M_{\rm min}}^{M_{\rm max}} dM \, \frac{dn}{dM} 
b_h(M) \tilde u(k|M) \right]^2P_{\rm lin}(k,z),
\label{eq:psdelta2}
\eea
where $\tilde u(k|M)$ is the Fourier transform of $\rho_{\rm m}(\mathbf x|M)/\bar \rho_{\rm m}$ ($\rho_{\rm m}$ and $\bar \rho_{\rm m}$ are respectively the matter density distribution and its cosmological average value), $P^{\rm lin}$ is the linear matter power spectrum and $b_{\rm h}$ is the linear bias (taken from the model of ref. \cite{Sheth:1999mn}).

To compute the 3D PS of galaxies, we need first to describe how galaxies populate halos. We employ the halo occupation distribution (HOD) formalism, following the approach described, e.g., in \cite{Zheng:2004id}. The occupation of galaxies is described in terms of central and satellite galaxies, $N_g=N_{\rm cen}+N_{\rm sat}$, as follows:
\bea
\langle N_{\rm cen}(M)\rangle &=& \frac{1}{2}\left[1+{\rm erf}\left(\frac{\log M-\log M_{\rm th}}{\sigma_{\rm logM}}\right)\right] \label{eq:HOD1}\\
\langle N_{\rm sat}(M)\rangle &=& \frac{1}{2}\left[1+{\rm erf}\left(\frac{\log M-\log 2 M_{\rm th}}{\sigma_{\rm logM}}\right)\right]\left( \frac{M}{M_1} \right)^\alpha \label{eq:HOD2}
\eea
$M_{\rm th}$ denotes the approximate halo mass required to populate the halo with the considered type of galaxies. The transition from 0 to 1 central galaxy is modeled by means of Eq.~\ref{eq:HOD1}, and set by the width $\sigma_{\rm LogM}$.
The satellite occupation is described by including a power law with index $\alpha$ and normalization set by the mass $M_1$.
The values of the four HOD parameters are taken from Ref.~\cite{Helgason_2012} for low-z galaxies and from Ref.~\cite{Cooray_2012} for high-z galaxies. In the first (latter) case we have $M_{\rm th}= 10^9\,(10^6) \,M_\odot$, $\sigma_{\rm LogM}=0.2\,(0.3)$, $\alpha=1.0\,(1.5)$, and $M_1=5\times 10^{10} \,(1.5\times 10^7)\,M_\odot$.

The 3D PS is then:
\bea
P^{gg}_{1h}(k,z) &=& \int_{M_{\rm min}}^{M_{\rm max}} dM \, \frac{dn}{dM} \frac{2\langle N_{\rm sat}\rangle\,\langle N_{\rm cen}\rangle\, \tilde v(k|M)+\langle N_{\rm sat}\rangle^2\, \tilde v^2(k|M) }{\bar n_g^2}\nonumber \\
P^{gg}_{2h}(k,z) &=& \left[ \int_{M_{\rm min}}^{M_{\rm max}} dM \, \frac{dn}{dM} \,\frac{\langle N_{g}\rangle}{\bar n_g}\,
b_h(M) \tilde v(k|M) \right]^2 P_{\rm lin}(k,z)\;,
\label{eq:psgal}
\eea
with $\tilde v(k|M)=\tilde u(k|M)\,\bar \rho_{\rm m}/M$.

In the case of IHL, we assume the emission profile to follow the host-halo dark matter profile and model the 3D PS as in Ref.~\cite{cooray2012measurement}:
\bea
P^{II}_{1h}(k,z) &=& \int_{M_{\rm min}^I}^{M_{\rm max}^I} dM \, \frac{dn}{dM} \left(\frac{L_{IHL,\lambda}(M,z)}{4\pi\,(1+z)\,W_I(\lambda,z)}\right)^2 \tilde v^2(k|M)\nonumber \\
P^{II}_{2h}(k,z) &=& \left[ \int_{M_{\rm min}^I}^{M_{\rm max}^I} dM \, \frac{dn}{dM} \,\frac{L_{IHL,\lambda}(M,z)}{4\pi\,(1+z)\,W_I(\lambda,z)}\,
b_h(M) \tilde v(k|M) \right]^2 P_{\rm lin}(k,z),
\label{eq:psgal}
\eea

Let us now move the cross-correlation PS between different source fields. Such contributions were neglected in Ref.~\cite{Cooray_2012}, but they can be relevant. 
The 3D PS of cross-correlation can be written as:
\bea
 P_{i,j}^{1h}(k,z) &=& \int_{{\rm max}(M_{\rm min}^i,M_{\rm min}^j)}^{{\rm min}(M_{\rm max}^i,M_{\rm max}^j)} dM\ \frac{dn}{dM} \tilde f_i(k|M)\, \tilde f_j(k|M) \label{eq:Pgen}\\
 P_{i,j}^{2h}(k,z) &=& \left[\int_{M_{\rm min}^i}^{M_{\rm max}^i} dM\,\frac{dn}{dM} b_h(M)  \,\tilde f_i(k|M) \right] \left[\int_{M_{\rm min}^j}^{M_{\rm max}^j} dM\,\frac{dn}{dM} b_h(M)\, \tilde f_j(k|M) \right]\,P^{\rm lin}(k)\;,\nonumber
\\
 {\rm with}&\;& \tilde f_a=\tilde u\;\;\;,\;\;\; \ \tilde f_g=\frac{\langle N_{g}\,\rangle}{\bar n_{g}}\tilde v\;\;\;,\;\;\; \ \tilde f_I=\frac{L_{IHL,\lambda}(M,z)}{4\pi\,(1+z)\,W_I(\lambda,z)}\,\tilde v\;.
\eea

\section{Experiments and data}
\label{sec:expdata}

The angular fluctuations of the near-IR background have been observed in many different experiments~\cite{2018RvMP...90b5006K, Kashlinsky:2005di}. 
In this work we selected data with the goal of covering the largest possible frequency range with the most constraining observations.
We analysed the measurements of the Hubble Space Telescope (HST)~\cite{Windhorst:2010ib} at wavelengths $\lambda= 0.606, 0.775, 0.85, 1.25, 1.6$ $\mu {\rm m}$\cite{Mitchell_Wynne_2015} and the observations by Spitzer Deep Wide-Field Survey (SDWFS) at wavelengths $\lambda= 3.6, 4.6$ $\mu {\rm m}$\cite{Kashlinsky:2012zz}. 

We also considered the data from the AKARI satellite\cite{Seo:2015fga} at $\lambda = 2.4, 3.2$ $\mu {\rm m}$ and from the Cosmic Infrared Background Experiment (CIBER)\cite{Bock:2012fw} $\lambda = 1.1, 1.6$ $\mu {\rm m}$.
We found CIBER data to be less constraining than HST data in the same wavelength range. Even though AKARI data do not overlap with the HST and Spitzer ones, they also do not provide any improvement in the ALP bounds. This is because at the AKARI depth, 
a contribution from ALP decay
compatible with the HST and Spitzer measurements would be a subdominant component of the total emission. We found AKARI data to be consistent with shot noise and known galaxy populations, in agreement with Ref. \cite{Helgason:2016xoc}.
Therefore, 
for the sake of simplicity, we will 
show our results only for HST and Spitzer. 

Being the data-sets quite separated in wavelength, they provide, when combined together, stringent constraints on the spectrum of the excess, challenging the ALP decay explanation. 

In the following, we will perform a statistical analysis to fit the predictions of the models discussed in~\Sec{sec:Models} to the experimental
measurements.
First, we will consider simultaneously the HST and Spitzer data-sets.
Then, we will also fit them separately, 
in order to understand the individual constraining power (we will refer to these cases as HST-only and Spitzer-only).
Furthermore, we will
explore both the cases where the IHL emission is included or neglected in the analysis. This will allow us to examine the relevance of this component, and its interplay with the emission from ALP decays to explain the near-IR excess.

\begin{table}
  \centering
  \renewcommand{\arraystretch}{2}
\begin{tabular}[7pt]{c | c | c | c}
        \hline\hline
        Parameters & ALPs + IHL  & ALPs-only & Prior ranges\\
        \hline
        $m_a$ & $2.85^{+0.06}_{-0.09}$ eV &  $2.45^{+0.09}_{-0.07}$ eV & $ (0, 10) $ eV\\
        $\tau_a$ & $(4.10^{+0.43}_{-0.28})\times 10^{23}$ s & $(4.28^{+0.32}_{-0.22})\times 10^{23}$ s & $(10^{20} , 10^{28})$ s \\
        $A_{IHL}$ & $ 0.168^{+0.040}_{-0.039}$ & -- & $(0, 10^3)$ \\
        $A_{06}$ & $(1.83^{+0.18}_{-0.16}) \times 10^3$ & $(1.89^{+0.16}_{-0.16}) \times 10^3 $ & $(0, 10^7)$\\
        $A_{07}$ & $ (3.40^{+0.30}_{-0.35}) \times 10^3 $ &  $ (3.58^{+0.30}_{-0.34})\times 10^3 $ & $(0, 10^7)$ \\ $A_{08}$ & $ (3.61^{+0.30}_{-0.26})\times 10^3 $ & $( 3.85^{+0.26}_{-0.30}) \times 10^3$ & $(0, 10^7)$ \\ $A_{12}$ & $( 1.08^{+0.59}_{-0.48}) \times 10^4$ & $ (4.73^{+4.17}_{-3.29}) \times 10^3 $ & $(0, 10^7)$ \\ $A_{16} $ & $ (2.22^{+0.33}_{-0.38}) \times 10^4$ & $(1.43^{+0.37}_{-0.31}) \times 10^4 $ & $(0, 10^7)$ \\  $A_{36} $ & $ 17.2^{+3.6}_{-3.8}$ & $ 19.2^{+3.7}_{-4.1}$ & $(0, 10^3)$\\
        $A_{45}$ & $ 25.9^{+2.4}_{-2.8}$ & $28.5^{+2.7}_{-2.2}$ & $(0, 10^3)$ \\ $\chi^2(\chi^2_{\text{reduced}})$ & $192 (1.68)$ & $217 (1.89)$ & -- \\
        \hline\hline
    \end{tabular}
    \caption{\textbf{Best fit HST + Spitzer.} Best fit values with $1\sigma$ intervals for the combined analysis HST + Spitzer, with or without the IHL contribution. We refer the Reader to Appendix~\ref{app:fit} for the best fit values also for HST-only and Spitzer-only analysis. } \label{tab:param_HSTSpitzer}
    \label{tab:HSTSpitzer}
\end{table}

\section{Results}
\label{sec:results}
The angular power spectrum at a given wavelength $\lambda$ and multipole $\ell$ is obtained summing the various contributions that we described in~\Sec{sec:Models}:
\begin{equation}
    C_\ell^{\rm model}(\lambda) = C_\ell^{\rm shot}(\lambda) + C_\ell^{f}(\lambda) + \sum_{i,j}C_{\ell}^{(i,j)}(\lambda)\;,
    \label{eq: model}
\end{equation}
where $i,j=$ low-z  residual galaxies,  high-z faint galaxies, ALP decays, and the IHL emission.
We note here that the window function of high-z galaxies has essentially no overlap in $z$ with the window functions of the other components, and therefore all the cross-correlation terms associated to high-z galaxies are negligible, see Eq.~\ref{eq:Cl}.

Our goal is to compare the prediction of the model with the measurements discussed in~\Sec{sec:expdata}.
For this purpose we build a likelihood function $\mathcal{L} \propto \exp(-\chi^2/2)$ with the chi-square given by

\bea
\label{eq:chi2}
    \chi^2 &=& 
    \sum_{\rm exp}\, \chi^2_{\rm exp}\,~~~{\rm with}\\
    \chi^2_{\rm exp} &=&
    \sum_{\lambda,\ell}\,
    \frac{\Big(C_\ell^{\rm data}(\lambda)- C_\ell^{\rm model}(\lambda)\Big)^2}{\sigma_\ell^2},
\eea    

where the sums run over the experimental data-sets considered, i.e. HST and Spitzer, the wavelength bands (5 for HST and 2 for Spitzer), and the multipoles.  
$C_\ell^{\rm data}(\lambda)$ is the experimental measurement of the angular power spectrum at a given wavelength and multipole, and $\sigma_\ell$ is the corresponding experimental uncertainty.
In total we have 125 data points, of which 70 from HST and 55 from Spitzer.

In the model fitting process, we explore the posterior distribution with a Markov
Chain Monte Carlo (MCMC) analysis employing the Metropolis-Hastings alghorithm~\cite{10.1093/biomet/57.1.97,Metropolis1953}, implemented through the Python ensemble sampler Emcee\cite{Foreman_Mackey_2013}. For our data-sets, $10^4$ samples are accumulated with 32 chains. We checked the chains show good acceptance rate and convergence. 
We notice that the parameter space is large and intricated, with different local minima, and therefore it is very important 
to scan it carefully.

In the next section we report the analysis of the combination HST + Spitzer; we will then specialize to the case in which we consider separately these two data-sets. We anticipate that the best fit parameters for the inclusive analysis are also close to the ones found for the fit of  the HST data. Instead, an analysis focused on the Spitzer measurements points to a different region of the ALP parameter space.

\begin{figure}[H]
\centering
\includegraphics[width=0.45\textwidth]{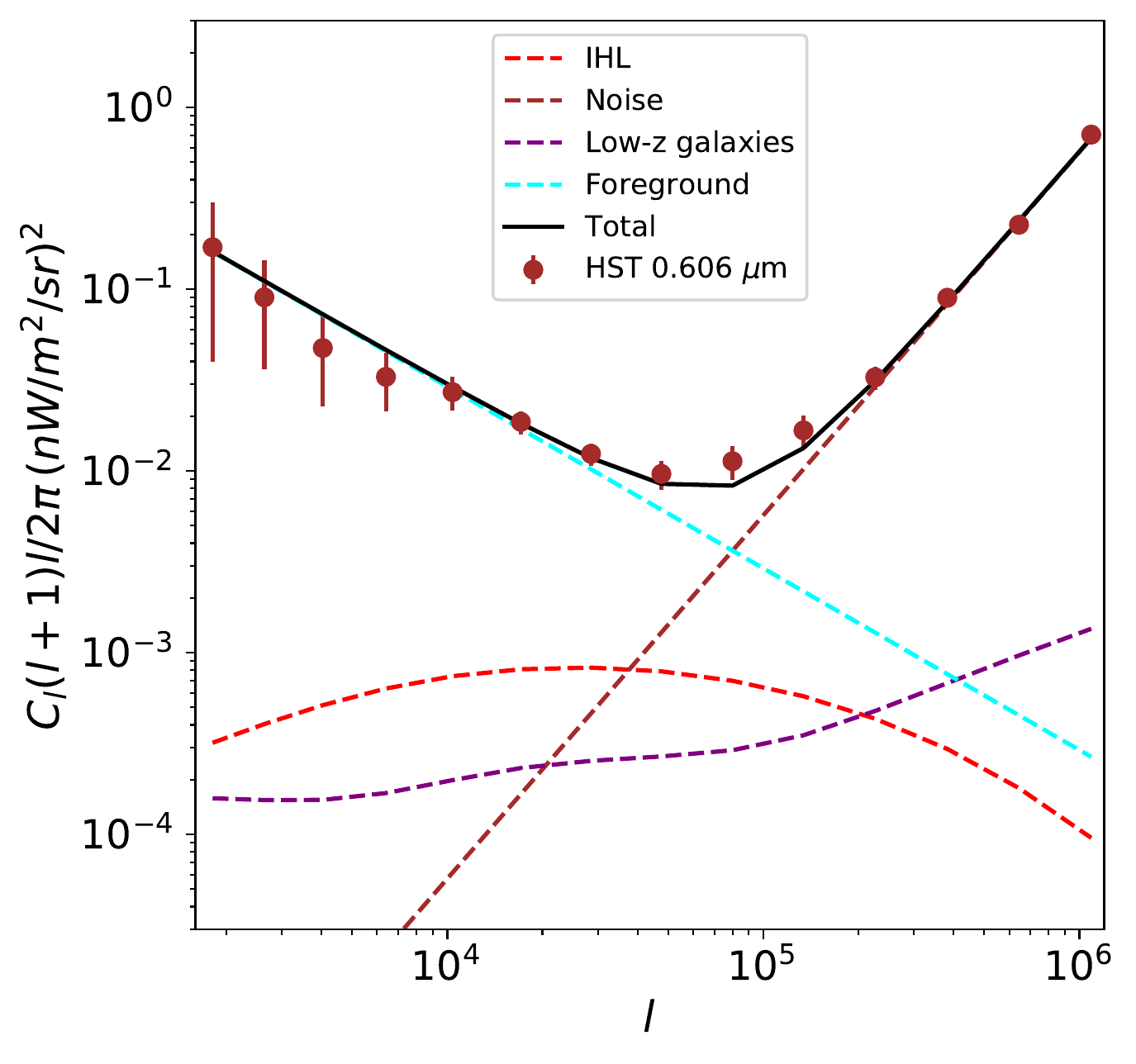}\hfill
\includegraphics[width=0.45\textwidth]{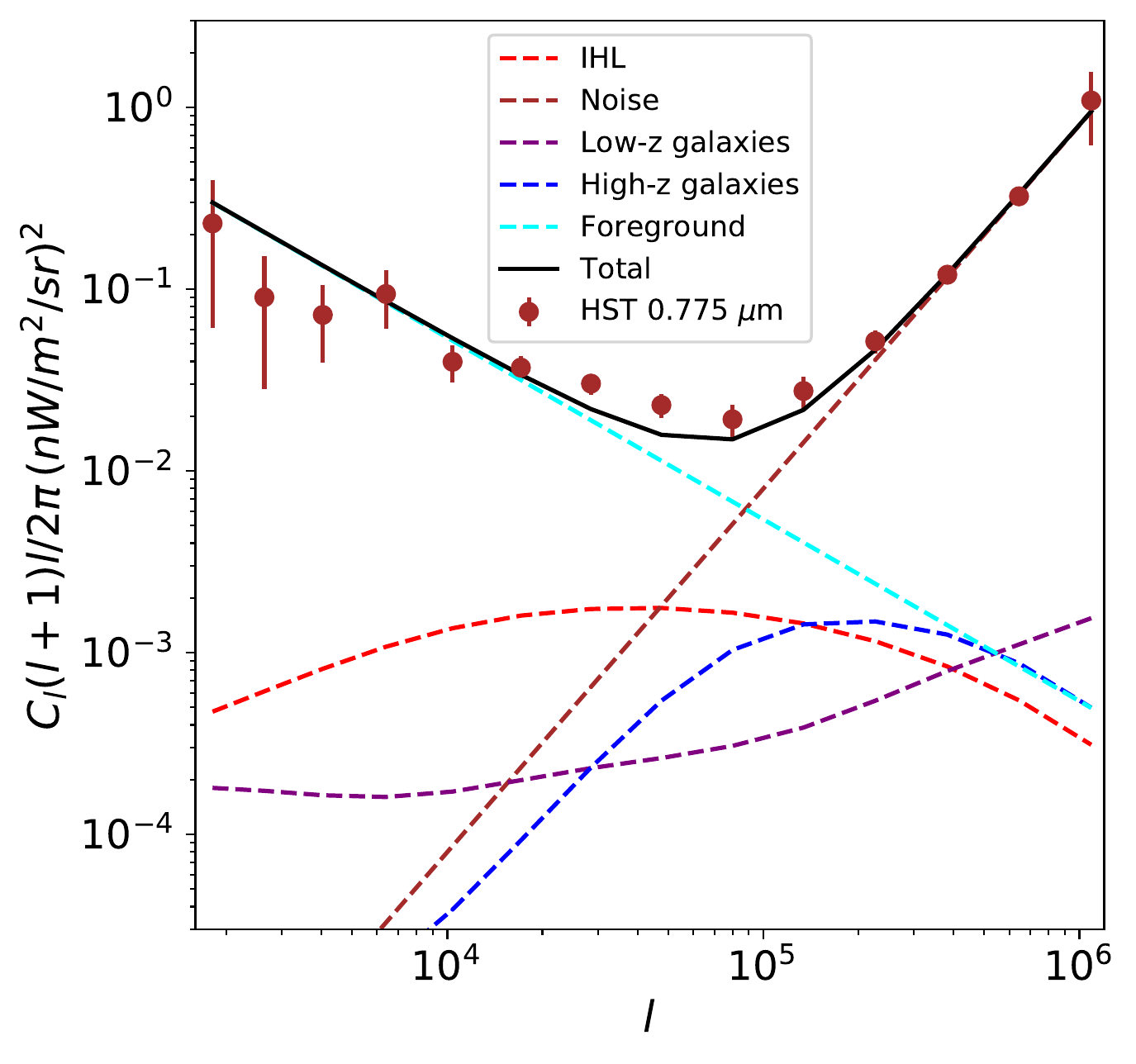}
\includegraphics[width=0.45\textwidth]{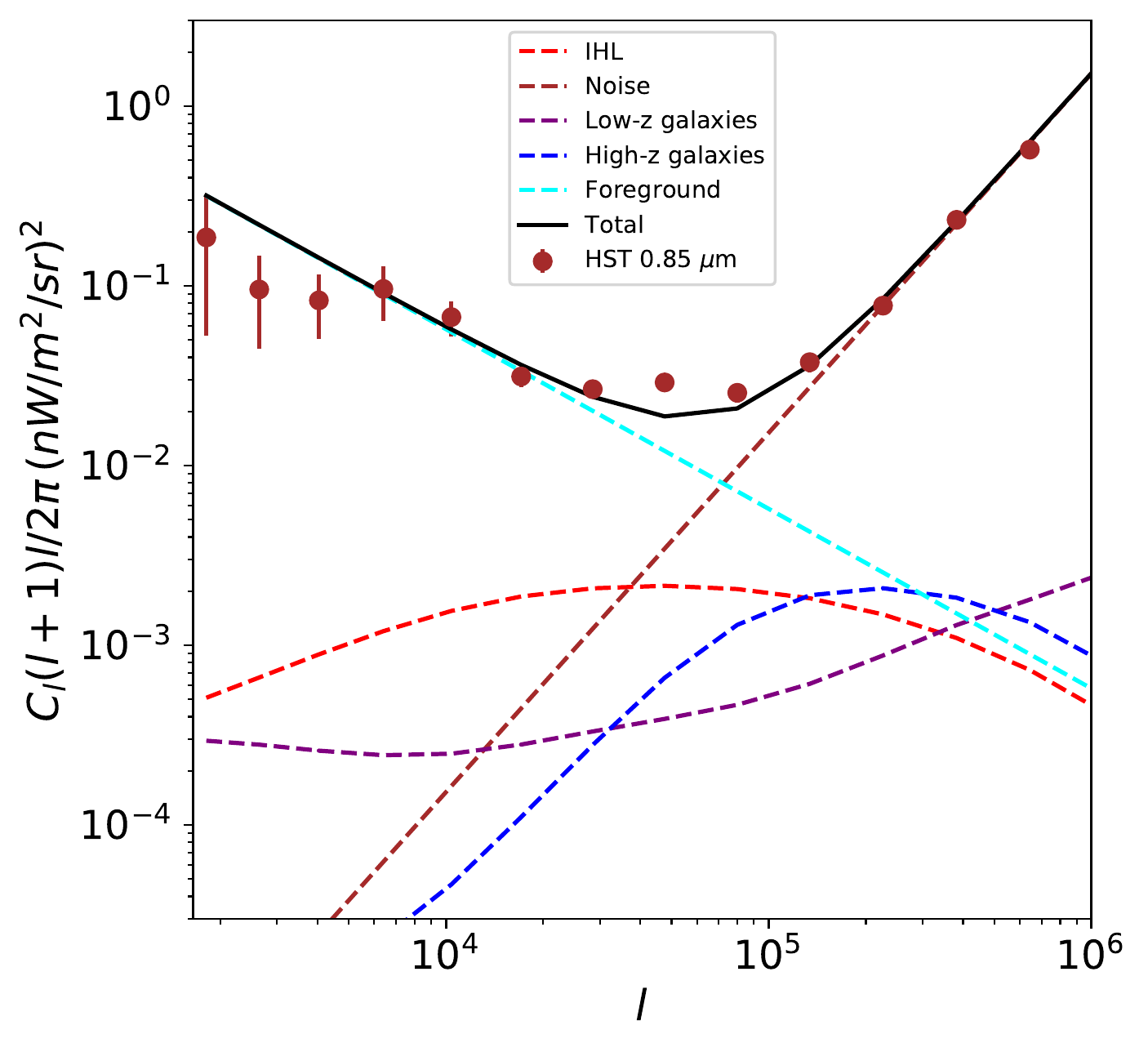}
\caption{\textbf{High-frequency HST bands for the combined analysis HST + Spitzer}. We show the different contributions (see labels) in the high-frequency HST bands for the best-fit model. ALP dark matter does not contribute to the fit, since the best-fit mass is smaller than twice the minimum frequency of these bands. The main contributions come from foregrounds, shot noise, IHL and low-z galaxies.}
\label{fig:fit_HSTlow1}
\end{figure}

\begin{figure}[H]
\centering
\includegraphics[width=0.5\textwidth]{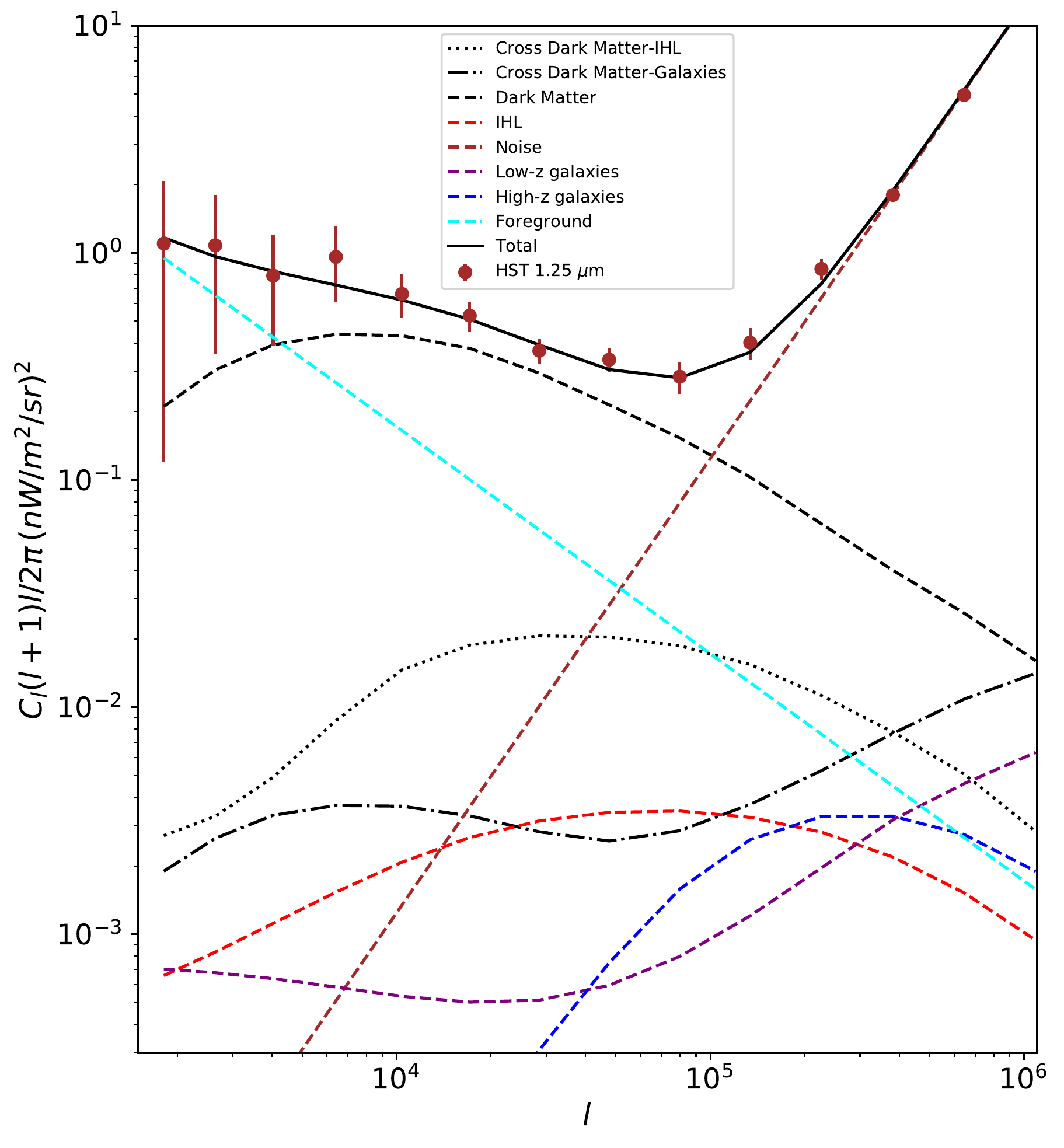}\hfill
\includegraphics[width=0.5\textwidth]{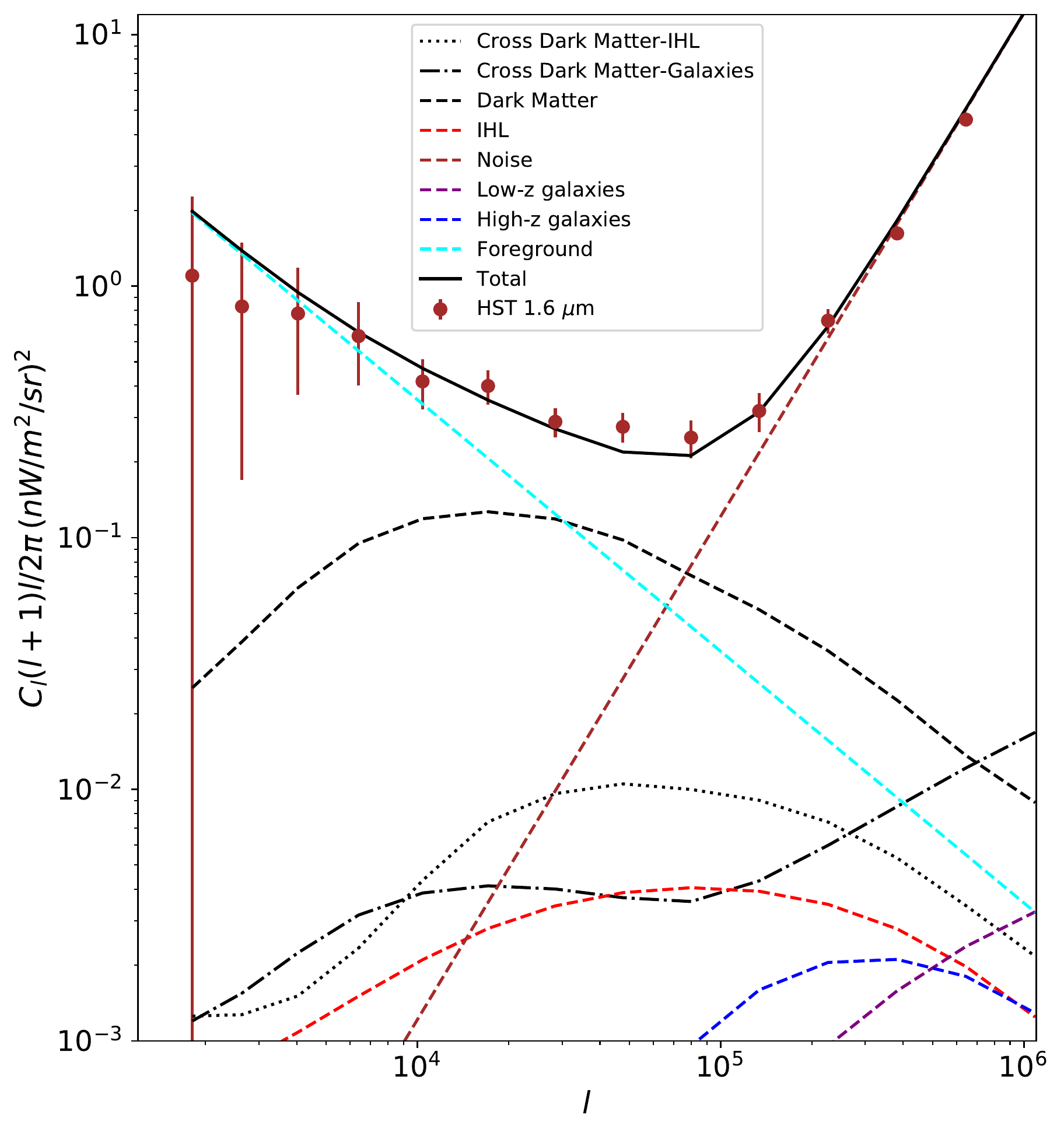}
\caption{\textbf{Low-frequency HST bands for the combined analysis HST + Spitzer}. We show the different contributions (see labels) in the low-frequency HST band for the best-fit model.} 
\label{fig:fit_HSThigh}
\end{figure}

\subsection{HST and Spitzer}

\begin{figure}[t]
\centering
\includegraphics[width=0.5\textwidth]{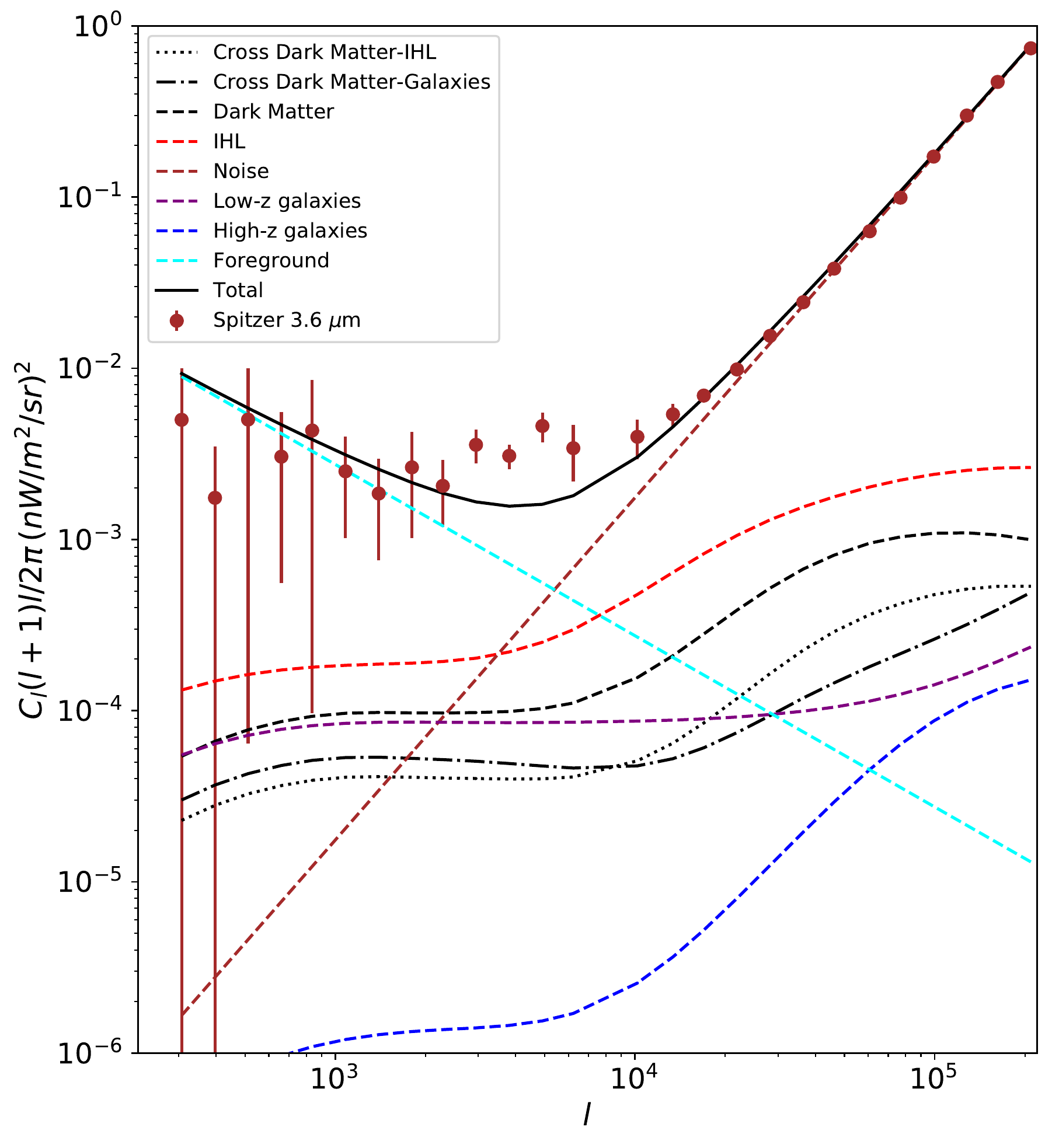}\hfill
\includegraphics[width=0.5\textwidth]{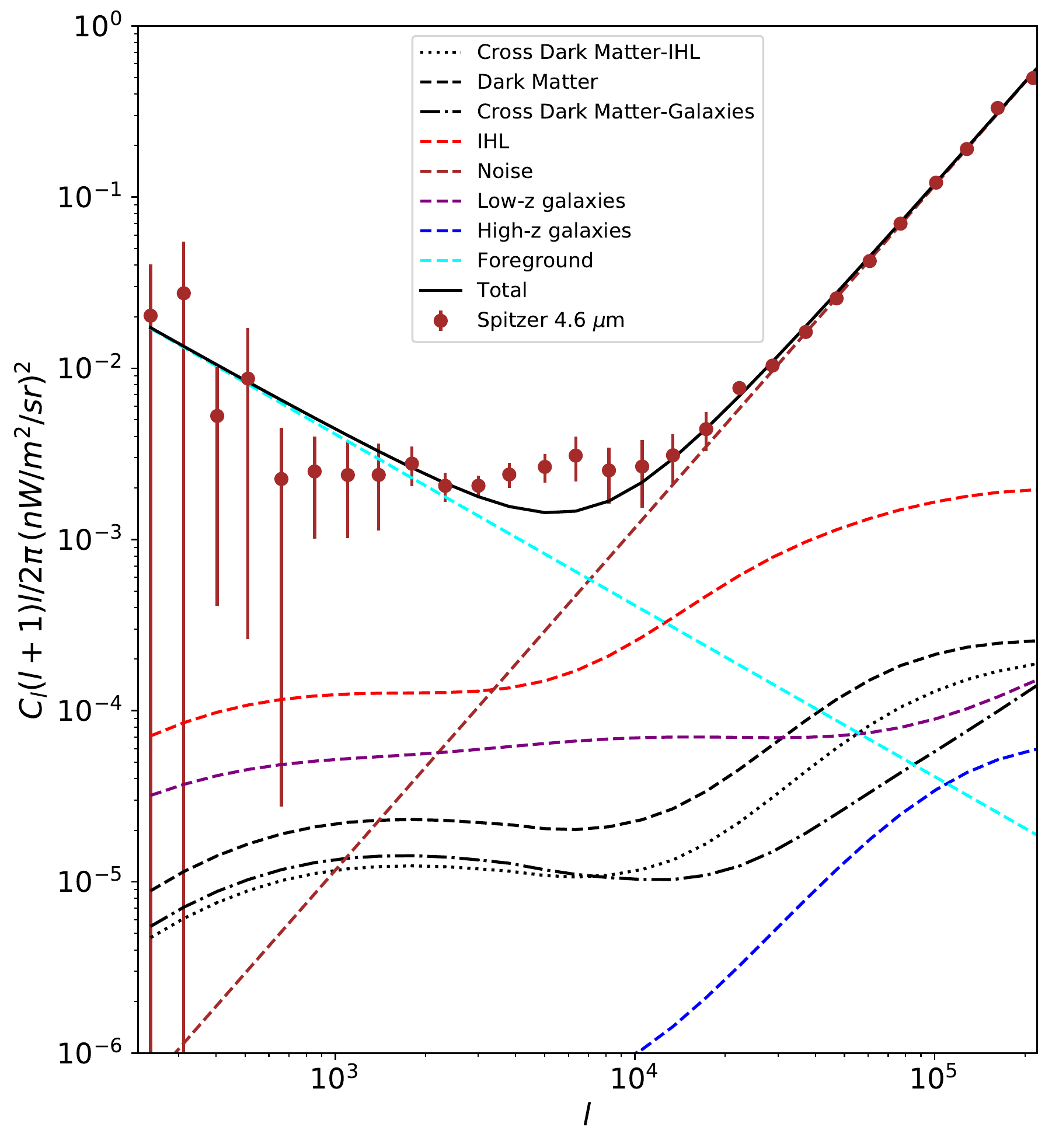}
\caption{\textbf{Spitzer bands for the combined analysis HST + Spitzer}. 
As in Figs.~\ref{fig:fit_HSThigh} and \ref{fig:fit_HSTlow1} but for the Spitzer wavelength bands.
} 
\label{fig:fit_Spitzer}
\end{figure}

The total chi-square defining the likelihood function in Eq.~\ref{eq:chi2} is given by 
\begin{equation}
    \chi^2 = \chi^2_{\rm HST} + \chi^2_{\rm Spitzer},
\end{equation}
As we already mentioned,
combining the two data-sets,
we have $N= 125$ datapoints. The number of free parameters is 10 (9) for the case with (without) the IHL contribution. The free parameters are the ALP mass and lifetime, $m_a, \tau_a$, the IHL amplitude $A_{\rm IHL}$ and the foregrounds amplitude factors $A^f$, one for each of the considered band (therefore 7 in this case). We stress that we also had the shot noise amplitude factors as free parameters. However, as already anticipated, we fit them in each band before running the MCMC for two reasons: first, this should not interfere with the actual fitting procedure, given that the contributions from ALPs and the IHL are relevant at much larger angular scales; second, given the complexity of each 
term, we found numerically much more convenient to first fit the shot noise in order to determine a conservative limiting magnitude that specifies the 
emission from low-z galaxies.
For the same numerical reasons we also decided to keep fixed the parameters $f_* = 0.03$ and $M_{\rm min} = 10^6 M_{\odot}$ for the
high-z galaxies. 
This should not affect the fit because, as we will show, the contribution from
high-z galaxies 
is always subdominant. To confirm this we checked that adding two free-varying amplitudes $A_{\rm low-z}, A_{\rm high-z} \in [0.1, 10]$ in front of the low-z and high-z galaxies
components,
only mildly affects the $\chi^2$ analysis. In particular, 
the ALP parameters inferred from the fit
are not sensitive to a rescaling of the contribution from high-z galaxies,
while they can change by few percent if the normalization of the low-z
galaxy emission is left free, nevertheless without altering our conclusions on the axion parameter space. This was actually physically expected. It is in fact evident from Fig.~\ref{fig:fit_HSTlow1}-\ref{fig:fit_Spitzer} that the shapes of the angular power spectra of DM and galaxies are rather different. We therefore decided to proceed with the MCMC analysis without the inclusion of these extra free parameters.

The best-fit $\chi^2$ for the case ALPs + IHL (ALPs-only) is:
\begin{equation}
    \chi^2 = 192 \;\; (217), \, \, \chi^2_{\rm reduced} = 1.68 \;\; (1.89),
\end{equation}
with ALP best fit parameters being 

\begin{eqnarray}
    m_a = 2.847^{+0.057}_{-0.086} \, {\rm eV} \;\; \Big( 2.452^{+0.088}_{-0.074} \Big), \\
    g_{a\gamma\gamma} = 1.191^{+0.047}_{-0.054} \times 10^{-10}\, {\rm GeV}^{-1} \;\;\Big( 1.451^{+0.045}_{-0.061}\Big).
\end{eqnarray}
Best fit values and 1$\sigma$ intervals  for all the parameters of the fit are presented in Table~\ref{tab:HSTSpitzer}.

It is evident that the combination of the IHL and ALP signals helps in explaining the IR excess. Fitting the data with IHL only would make the fit significantly worst, with $\chi^2 \simeq 262$. 
In Fig.~\ref{fig:fit_HSTlow1} we show the various contributions in the high-frequency HST bands for the best fit case (ALPs + IHL). The ALP decay does not contribute in these plots since the maximal frequency of emission ($m_a/2$) for the best-fit case is smaller than the lowest frequency band.
We note that cross-correlation terms are not shown, since they are below the reported vertical scale in the plots. The fact they are subdominant is not surprising since in this plot they involve correlation terms between IHL (given by stars outside galaxies) and galaxies, which are expected to be small.

In the low-frequency HST band, Fig.~\ref{fig:fit_HSThigh}, ALPs play a leading role, contributing to provide a very good fit to the data. In this case the IHL is subleading, as well as the galaxies.
Note that cross terms (now including the cross-correlation of ALPs with IHL and low-z galaxies) are sizeable even though subdominant with respect to the ALPs auto-correlation term.

In Fig.~\ref{fig:fit_Spitzer}, we show the results of the fit for what concerns the Spitzer data-set. In this case both the IHL and ALP decay provide significant contributions. However the excess at intermediate scales is not reproduced in a fully satisfactorily way. 

\begin{figure}[H]
\centering
\includegraphics[width=7cm]{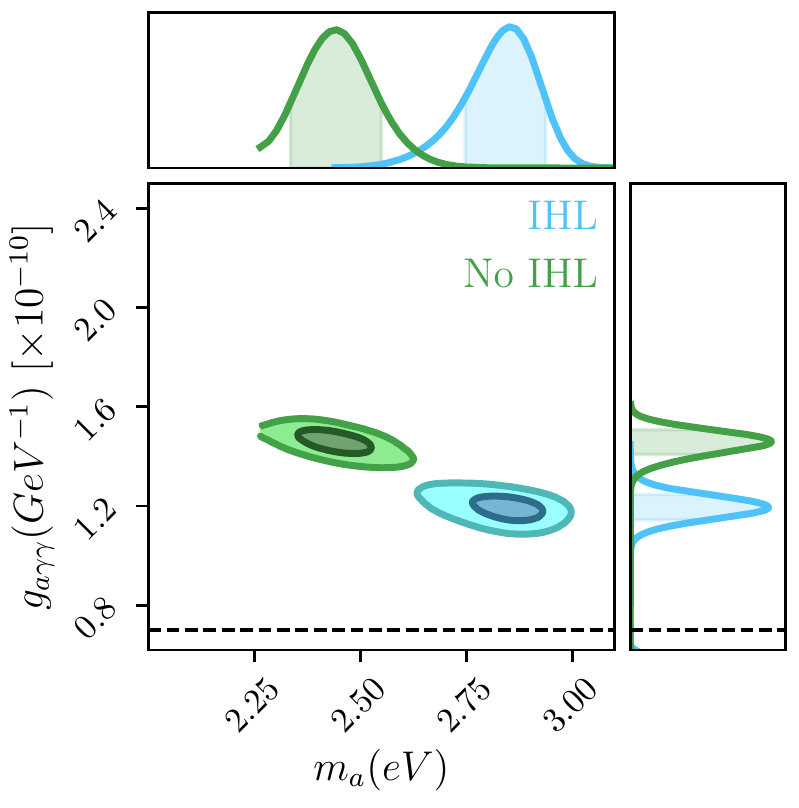}\caption{\textbf{ALP posterior distributions for the combined analysis HST + Spitzer}. $68\%$ and $95\%$ C.L regions in the ALP parameter plane for the posterior of the combined analysis HST + Spitzer. We show results for both the analysis with (light blue) and without (green) the IHL contribution. The horizontal black dashed line is the $95\%$C.L exclusion limit from Horizontal Branch (HB) stars~\cite{Ayala:2014pea}.} 
\label{HST_Spitzer}
\end{figure}

Finally, in Fig.~\ref{HST_Spitzer} we show the $68\%$ and $95\%$ C.L regions for marginalized posterior distribution in the ALP parameter space. The horizontal black dashed line is the $95\%$ C.L. exclusion limit from Horizontal Branch (HB) stars. We then see that the preferred regions of our fit are
in tension with stellar cooling bounds.

\subsection{HST only}
From the analysis of the previous section, we find that the combination of IHL and ALP decay partially explains the excess in the HST and Spitzer data, but the combined fit is not good, as one can see from the $\chi^2$ (which leads to a very small p-value).
It is therefore of interest to ask whether ALP dark matter may explain the excess only in some frequency bands. 
We start studying HST data, namely the high-frequency ones. 
The chi-square 
in Eq.~\ref{eq:chi2}
now reads

\begin{equation}
    \chi^2 = \chi^2_{\rm HST},
\end{equation}
where $\chi^2_{HST}$ includes the five HST frequency bands.

The ALP best fit parameters for the case ALPs + IHL (ALPs-only) are 
\begin{eqnarray}
    m_a = 2.498^{+0.223}_{-0.163} \, {\rm eV} \;\; \Big( 2.580^{+0.198}_{-0.184} \Big), \\
    g_{a\gamma\gamma} = 1.287^{+0.123}_{-0.116} \times 10^{-10} {\rm GeV}^{-1} \;\; \Big( 1.376^{+0.123}_{-0.102}\Big),
\end{eqnarray}
and the chi-squares are
\begin{equation}
    \chi^2 = 59.2 \, (116.2), \, \, \chi^2_{\rm reduced} \sim 0.97 \, (1.9).
\end{equation}

The combination of IHL and ALP decay provides an excellent fit for the near-IR excess in the HST bands (p-value $=0.54$). 
As already mentioned, we checked 
that the inclusion of a free amplitude for the low-z galaxies does not modify the best-fit point and the chi-square.

\begin{figure}[H]
\centering
\includegraphics[width=0.49\textwidth]{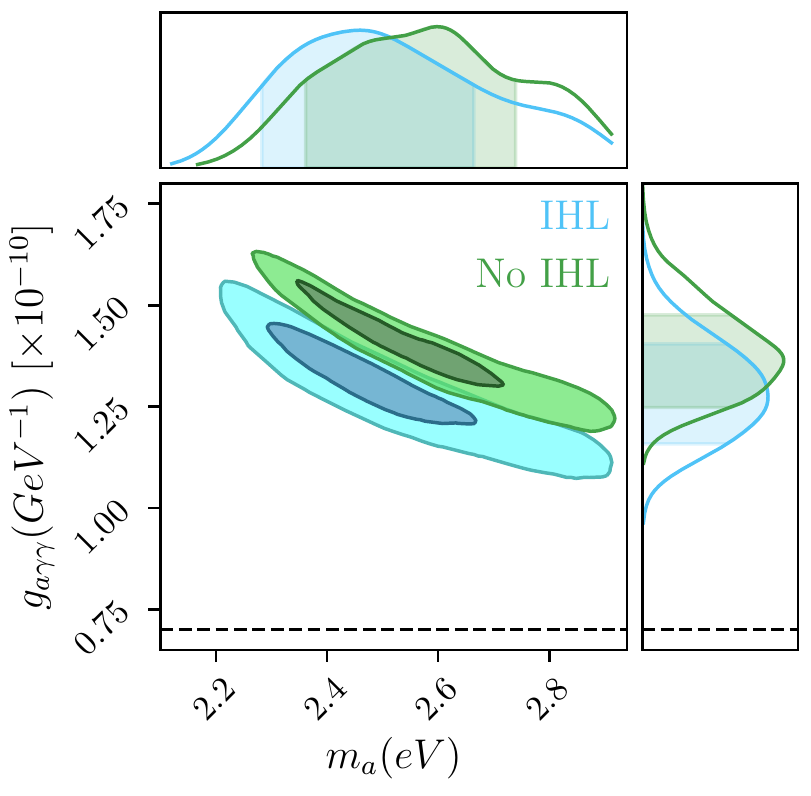}
\includegraphics[width=0.49\textwidth]{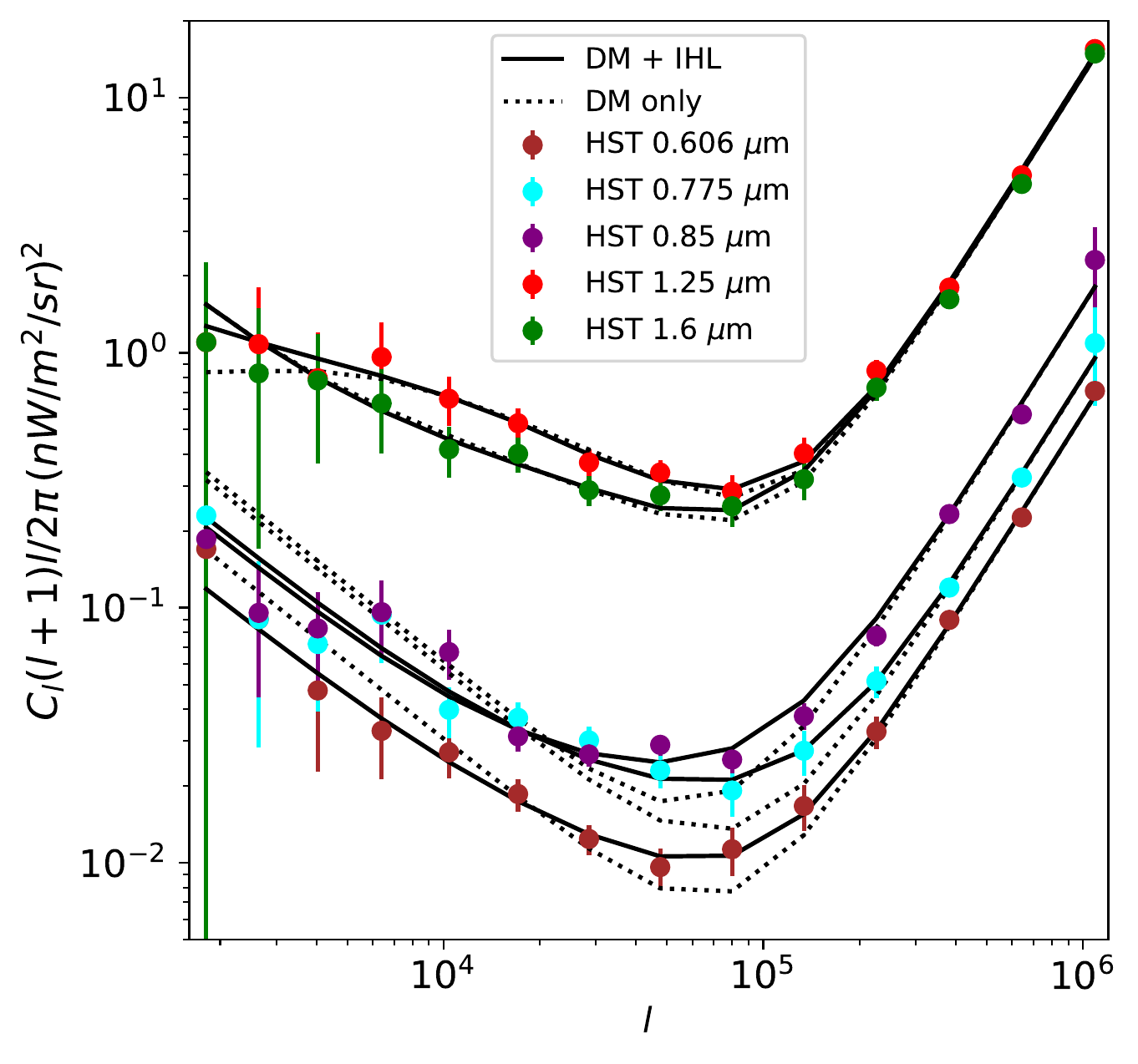}\hfill
\caption{\textbf{HST-only analysis:}\ \textbf{Left panel:} $68\%$ and $95\%$ C.L regions in the ALP parameter plane for the posterior of the analysis of HST-only data. We show results for both the analysis with (light blue) and without (green) IHL contribution. The horizontal black dashed line is the $95\%$C.L exclusion limit from Horizontal Branch (HB) stars. \textbf{Right panel:} data and best-fit model in the different HST frequency bands, both with (solid black) and without (dotted black) the IHL contribution. 
The combination of IHL and ALP decays provides an excellent fit to HST data (see the text).} \label{HST}
\end{figure}

We show the marginalized posterior in Fig.~\ref{HST}. The preferred regions for the ALP parameters are not dramatically changed with respect to the combined analysis HST+ Spitzer. 
This was expected since HST data are more constraining than the Spitzer ones (larger number of points and smaller relative errors).
This can be noted also from Fig.~\ref{fig:fit_HSTlow1}-\ref{fig:fit_HSThigh} where ALPs plus IHL nicely fit the HST data excess, while in the Spitzer data, Fig.~\ref{fig:fit_Spitzer}, an excess somewhat remains and the main contributors to the APS are foregrounds and shot noise, also at intermediate scales. 

As for the combined analysis, the regions in Fig.~\ref{HST} are
in tension with the constraints from HB stars.

\subsection{Spitzer only}

Finally, we consider only the low-frequency data from Spitzer. In this case the best-fit in the ALP parameter space changes substantially from the previous analyses, pointing to lower ALP masses and a larger lifetime (thus even more in tension with the bounds from HB stars). We remind that lower is the frequency of observation, lower the ALP mass can be, still producing an observable radiation from the ALP decay.
Analogously to the other cases we have
\begin{equation}
    \chi^2 = \chi^2_{\rm Spitzer},
\end{equation}
where $\chi^2_{\rm Spitzer}$ 
includes the two Spitzer frequency bands. 
From our statistical analysis we find the following ALP best fit parameters:
\begin{eqnarray}
    m_a = 0.825^{+0.036}_{-0.032} \, \text{\rm eV}, \\
    g_{a\gamma\gamma} = 1.721^{+0.007}_{-0.065} \times 10^{-10} \text{\rm GeV}^{-1},
\end{eqnarray}
and
\begin{equation}
    \chi^2 = 26.2 \, \,, \chi^2_{\rm reduced} \sim 0.6\;.
\end{equation}

\begin{figure}[H]
\centering
\includegraphics[width=0.44\textwidth]{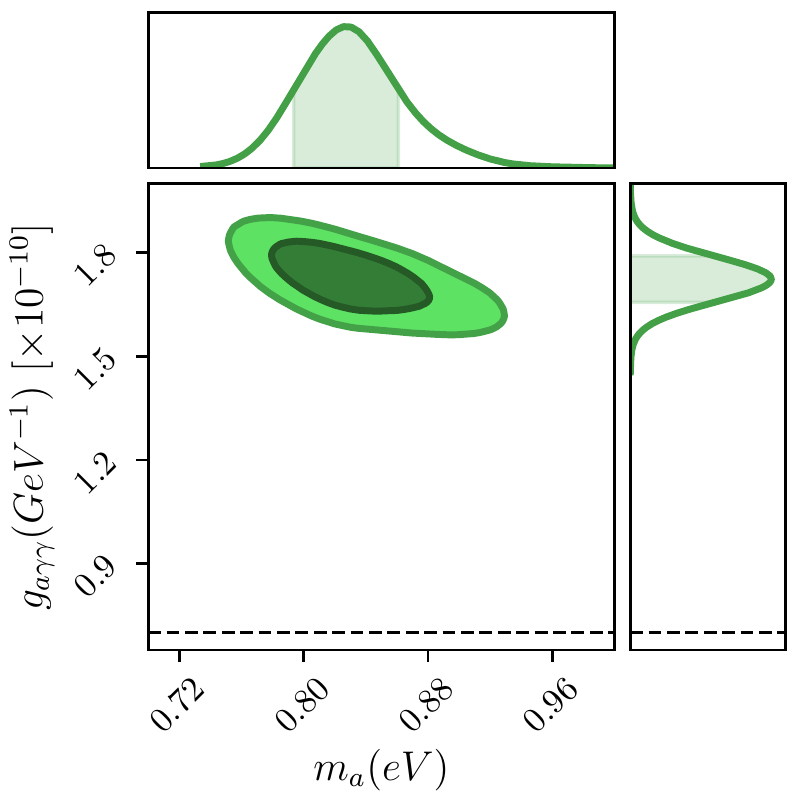}
\includegraphics[width=0.46\textwidth]{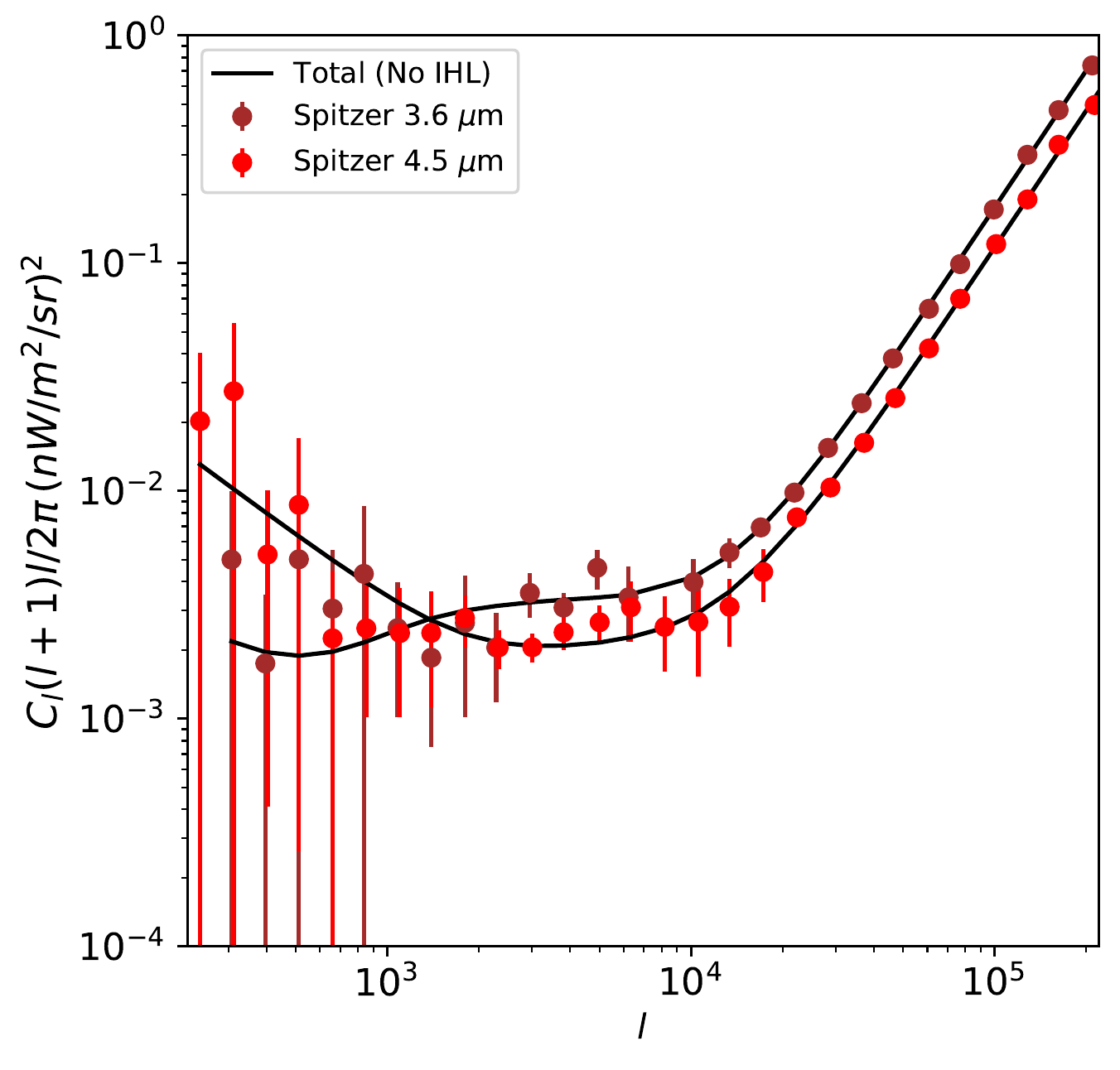}
\caption{\textbf{Spitzer-only analysis:} \textbf{Left panel:} $68\%$ and $95\%$ C.L regions in the ALP parameter plane for the posterior of the analysis of Spitzer-only data. We show the results for the ALPs-only analysis; in this case the addition of the IHL contribution does not produce any relevant change and the normalization of the IHL preferred by the fit is compatible with zero.
The horizontal black dashed line is the $95\%$C.L exclusion limit from Horizontal Branch (HB) stars. \textbf{Right Panel:} data and best-fit model for the Spitzer bands.
} 
\label{fig:SpitzerOnly}
\end{figure}

The value of the $\chi^2$ indicates that our model is slightly overfitting the data. This suggests there is no need for two extra components (ALPs and IHL) and one can be sufficient.
In fact, we find that the addition of the IHL does not help to fit Spitzer data, i.e., the $\chi^2$ is basically unchanged when including the IHL contribution, with the best fit value of $A_{IHL}$ (the amplitude of the IHL contribution) being always very small and compatible with zero within the errors.
Therefore, in the Spitzer-only case, we do not distinguish between the cases with and without the IHL contribution. 
We show the results of our MCMC in Fig.~\ref{fig:SpitzerOnly} (left panel), together with the best-fit APS in the two bands (right panel). 

Finally, before moving to our conclusions, let us explain the physical reason for the preference of a DM component in the fits to the various datasets that we have considered.
As clear from Figs.~\ref{fig:fit_HSTlow1} and~\ref{fig:fit_HSThigh}, an excess appears at intermediate scales. Galaxies tend to fail in explaining those scales since they show more correlation either at larger scales, due to the two-halo term (i.e., correlation between different galaxies), or at smaller scales (from the one-halo term) since their average size is small.
On the contrary, the one-halo of DM and IHL contributions can come from larger scales since they have weight functions peaked at very low redshift, thus corresponding to larger objects (in angular scales). Moreover, in the case of DM, the signal is mainly produced by large halos. Spitzer wavelengths correspond to larger redshift, so the power at intermediate scales is suppressed (halos are smaller in angular units) and it is more difficult to account for the excess. The fit improves in the case of Spitzer-only, since one can get a different best-fit DM mass, in particular, lower masses, which means that in this case the Spitzer wavelengths are produced at lower z than in the case of HST+Spitzer.

\section{Conclusions}
\label{sec:conclusions}

In this paper we have studied the
decay of ALP dark matter into two photons
as a potential origin of the anisotropy of the near-IR background intensity. 
Our analysis shows that ALPs 
decay can provide, together with IHL, an excellent fit to HST data and Spitzer data
separately.
Instead, when the two data-sets are simultaneously considered, an excess in the near-IR clustering data still remains, leading to a significantly poorer fit.
The preferred ALP parameters range at 95\% C.L. in the combined analysis is (see Fig.~\ref{HST_Spitzer})
\begin{equation}
    m_a \simeq (2.3 \div 3.0) \, {\rm eV}, \, \, \,
    g_{a\gamma} \simeq (1.1 \div 1.6)\times 10^{-10} \, {\rm GeV}^{-1}.
\end{equation}
This ALP region is in tension with HB stars cooling data~\cite{Ayala:2014pea}, which therefore disfavour an interpretation of the near-IR excess in terms of ALP decay.

A way to possibly evade stellar constraints involves ALPs with properties that depend on the environment, allowing for a suppressed production in stars. Examples of such models can be found in~\cite{Bloch:2020uzh, Masso:2005ym, Jaeckel_2007}.

We also notice that our best fit regions point to slightly smaller masses than  what was found in Ref.~\cite{Gong:2015hke}. This may be due to a different procedure for the scan of the parameter space. Using the same datasets and emission sources of Ref.~\cite{Gong:2015hke} (without \emph{e.g} the inclusion of the IHL and cross-terms), we found a local minimum in the parameter space which is located very close to the best fit point reported in Ref.~\cite{Gong:2015hke}, with $m_a \simeq 4$ eV, but our $\chi^2$ for this model is always substantially larger than the global one we have found for $m_a \simeq (2.4 \div 2.9)$ eV. Incidentally, the best fit point of Ref.~\cite{Gong:2015hke} seems now to be excluded by the recent analysis of MUSE spectroscopic observations of the dwarf spheroidal galaxy Leo T~\cite{Regis:2020fhw}.

Finally, it is worth mentioning the existence of a cross-correlation signal between the cosmic infrared and the cosmic X-ray backgrounds~\cite{Li_2018, Mitchell-Wynne:2016hgd, Cappelluti:2017nww, Cappelluti_2013}. This signal can further constrain ALPs and it has been often interpreted as an imprint from early accreting black holes. Here we decided to err on the conservative side and consider the near-IR signal only. Given that stellar cooling data and spectroscopic observations already challenge this interpretation, we did not add X-ray data. However, it may be of interest to consider the cross power spectra between near-IR and X-rays for ALP models which can avoid stellar cooling bounds. We leave this study for future work.

\subsubsection*{Acknowledgments}

We thank C. Evoli, M. Giannotti and S. Mishra-Sharma for discussions.
A.C. acknowledges support from the Israel Science Foundation (Grant No. 1302/19), the US-Israeli BSF (grant 2018236) and the German Israeli GIF (grant I-2524-303.7). A.C. acknowledges also hospitality and support from the MPP of Munich. 
M.R. acknowledges support by ``Deciphering the high-energy sky via cross correlation'' funded by the agreement ASI-INAF n. 2017-14-H.0. 
M.R. acknowledge funding from the PRIN research grant ``From  Darklight  to  Dark  Matter: understanding the galaxy/matter connection to measure the Universe'' No. 20179P3PKJ funded by MIUR.
M.T. acknowledges support from the INFN grant LINDARK.
N.F. acknowledges funding from the research grant The Anisotropic Dark Universe, No. CSTO161409, funded by Compagnia di Sanpaolo and University of Torino.
N. F. and M.T. acknowledge funding from the PRIN research grant ``The Dark Universe: A Synergic Multimessenger Approach", No. 2017X7X85K funded by the Italian Ministry of Education, University and Research (MIUR).
N.F. and M.R. acknowledge funding from the ``Department of Excellence" grant awarded by the Italian Ministry of Education, University and Research (MIUR).
N.F., M.R. and M.T. acknowledge funding from the research grant TAsP (Theoretical Astroparticle Physics) funded by Istituto Nazionale di Fisica Nucleare (INFN).

\appendix

\section{High-z galaxies} \label{app:High_z}

Several processes contribute to the infrared background. Following Ref.~\cite{Cooray_2012} we consider: direct emission from stars, Lyman-$\alpha$ line, free-free, free-bound and two photon processes. 
The differential stellar luminosity at a frequency $\nu$ is computed assuming 
a Planckian spectrum truncated at $h\nu = 13.6$ eV:
\begin{equation}
L_{\nu}^* = 
\begin{cases}
     \pi S_{*} B_{\nu}(T^{\text{eff}}_*), & \text{if}\ h\nu < 13.6\, \text{eV} \\
      0, & \text{otherwise}
\end{cases},
\end{equation}
where $B_{\nu}(T)= \frac{2 h\nu^3/c^2}{e^{h\nu/T} -1}$,
and $S_*$ is the surface area of the star, $S_*=4\pi R_*^2,$ with $R_*$ the stellar radius.
In the following we consider metal-poor stars (Pop II stars, following~\cite{Fernandez2006}), which provide a larger contribution to the IR-background when compared with metal-free stars (Pop III stars).
Their effective temperature depends on the stellar mass via the fitting formula $\log_{10}(T^{\text{eff}}_*/K) = 3.92 +0.704 x -0.138 x^2$, where we have defined (here and below) $x \equiv \log_{10}(M_{*}/M_{\odot})$, with $M_{*}$ being the stellar mass and $M_{\odot}$ the solar mass.
The surface area of the star can be computed as: 
\begin{equation}
4 \pi R_*^2=\frac{L_*^{\text{bol}}}{\sigma\, T_*^{\text{eff}}}
\end{equation}
with an intrinsic bolometric luminosity given by $\log_{10}(L_*^{\text{bol}}/L_{\odot}) =  0.138+4.28 x -0.653 x^2$, with $L_{\odot}$ the Sun's luminosity.

The luminosity due to the Lyman-$\alpha$ emission is computed as
\begin{equation}
    L^{\text{Ly}\alpha}_{\nu} = h\nu_{\text{Ly}\alpha}(\epsilon^{\text{rec}}_{\text{Ly}\alpha}+\epsilon^{\text{coll}}_{\text{Ly}\alpha})\phi(\nu_{\text{Ly}\alpha}-\nu, z) V(M_*),
    \label{eq:Lya}
\end{equation}
where $\nu_{\text{Ly}\alpha}$ is the frequency of Lyman-$\alpha$ photons. The emission volume in the nebulae surrounding the star is
\begin{equation}
    V(M_*) = \frac{\bar{Q}_{HI}(M_*)}{n_e^{\text{neb}}n^{\text{neb}}_{\text{\text{HII}}}\alpha_B^{\text{rec}}},
\end{equation}
where $\alpha^{\text{rec}}_{B}$ is the hydrogen case B-recombination coefficient, $n_e^{\text{neb}}, \,n^{\text{neb}}_{\text{HII}}$ are the number density of electrons and HII in the nebulae, which we take to be $n_e^{\text{neb}} = n^{\text{neb}}_{\text{HII}} = 10^4\, \text{cm}^{-3}$. The time-averaged photoionization rate is instead given by $\log_{10}(\bar{Q}_{\text{HI}}(M_*) {\rm s}^{-1}) = 27.80 + 30.68 \, x -14.80 \, x^2 + 2.50 \, x^3 $. 
In addition to the contribution from the stellar nebulae, we also consider the emission from the IGM. This term, modeled as in~\cite{Cooray_2012}, contributes negligibly to the total luminosity.
In Eq.~\ref{eq:Lya}, $\epsilon^{\text{rec}}_{\text{Ly}\alpha}$ is the Lyman-$\alpha$ recombination emission rate per unit volume~\cite{Canta}, while $\epsilon^{\text{coll}}_{\text{Ly}\alpha}$ is the collisional emission rate per unit volume~\cite{Canta}. 
Detailed expressions for these quantities as well as for $\alpha^{\text{rec}}_{B}$ can be found in~\cite{Cooray_2012}.
The function $\phi(\nu_{\text{Ly}\alpha} - \nu, z)$ is the line profile~\cite{Santos2002}:
\begin{equation}
\phi(\nu_{\text{Ly}\alpha} - \nu, z) = 
\begin{cases}
     \nu_*(z)d\nu^{-2}\exp(-\nu_*(z)/d\nu), & \text{if}\ d\nu >0 \\
      0, & \text{otherwise}
\end{cases},
\end{equation}
where $d\nu \equiv \nu_{\text{Ly}\alpha} - \nu$ and $\nu_*(z)= 1.5\times 10^{11} \text{GHz}\, (1+z)^3 H_0/H(z)$, with $H$ being the Hubble rate.

For the free-free and free-bound emission we follow Ref.~\cite{Fernandez2006}. The form of the luminosity produced by these two contributions is the same:
\begin{equation}
    L_{\nu}^{\text{ff, fb}} = 4\pi j^{\text{ff, fb}}_{\nu}\, V(M_*),
\end{equation}
where the specific emission coefficient reads
\begin{equation}
    j^{\text{ff, fb}} = 5.44 \times 10^{-39}\, \frac{e^{-h \nu/T}}{\sqrt{T}}\, n_e^{\text{neb}}\, n_{\text{HII}}^{\text{neb}}\, g_{\text{eff}}^{\text{ff, fb}} \,\,[\text{erg}/\text{cm}^3/\text{s}/\text{Hz}/\text{sr}],
\end{equation}
with 
$T$ the gas temperature (which we set to $T = 3\cdot 10^4$ K~\cite{Cooray_2012}) and $g_{\text{eff}}^{\text{ff, fb}}$ the Gaunt factor which takes the form 
\begin{equation}
g_{\text{eff}}^{\text{ff, fb}} = 
\begin{cases}
     1.1, & \text{free-free} \\
      \frac{x_n e^{x_n}}{n} \, 1.05, & \text{free-bound.}
\end{cases}.
\end{equation}
In the above formula $x_n \equiv R_y/(T n^2)$, where $R_y = 1.1 \times 10^7 \text{m}^{-1}$ is the Rydberg constant. The integer number $n$ indicates the energy level determined by the frequency of the emitted photon; if $R_y/n'^2 < \nu < R_y /(n' -1)^2$, then $n = n'$.
We consider $n>2$ because photons from the $n=1$ transition are efficiently absorbed by other neutral hydrogen atoms.
As for the Lyman-$\alpha$ emission, a subdominant contribution is obtained from the IGM emission.

Finally, we also consider two-photon processes~\cite{Fernandez2006}:
\begin{equation}
    L_{\nu}^{2\gamma} = \frac{2h \nu}{\nu_{\text{Ly}\alpha}}P(\nu/\nu_{\text{Ly}\alpha})\,\epsilon_{2\gamma}\,V(M_*),
\end{equation}
where $\epsilon_{2\gamma} = n_e n_{\text{HII}} \alpha_B^{\text{rec}}(1- f^{\text{rec}}_{\text{Ly}\alpha})$ is the two-photon emission rate, while $P(\nu/\nu_{Ly\alpha})$ is the normalizd probability of generating one photon in the range $d\nu/\nu_{Ly\alpha}$ from a two photon decay 
\begin{equation}
 P(y) = 1.307 - 2.627(y-0.5)^2 + 2.563(y - 0.5)^4 - 51.69(y - 0.5)^6,
\end{equation}
with $y \equiv d\nu/ \nu_{Ly\alpha}$.

\begin{figure}[H]
\centering
\includegraphics[width=10cm]{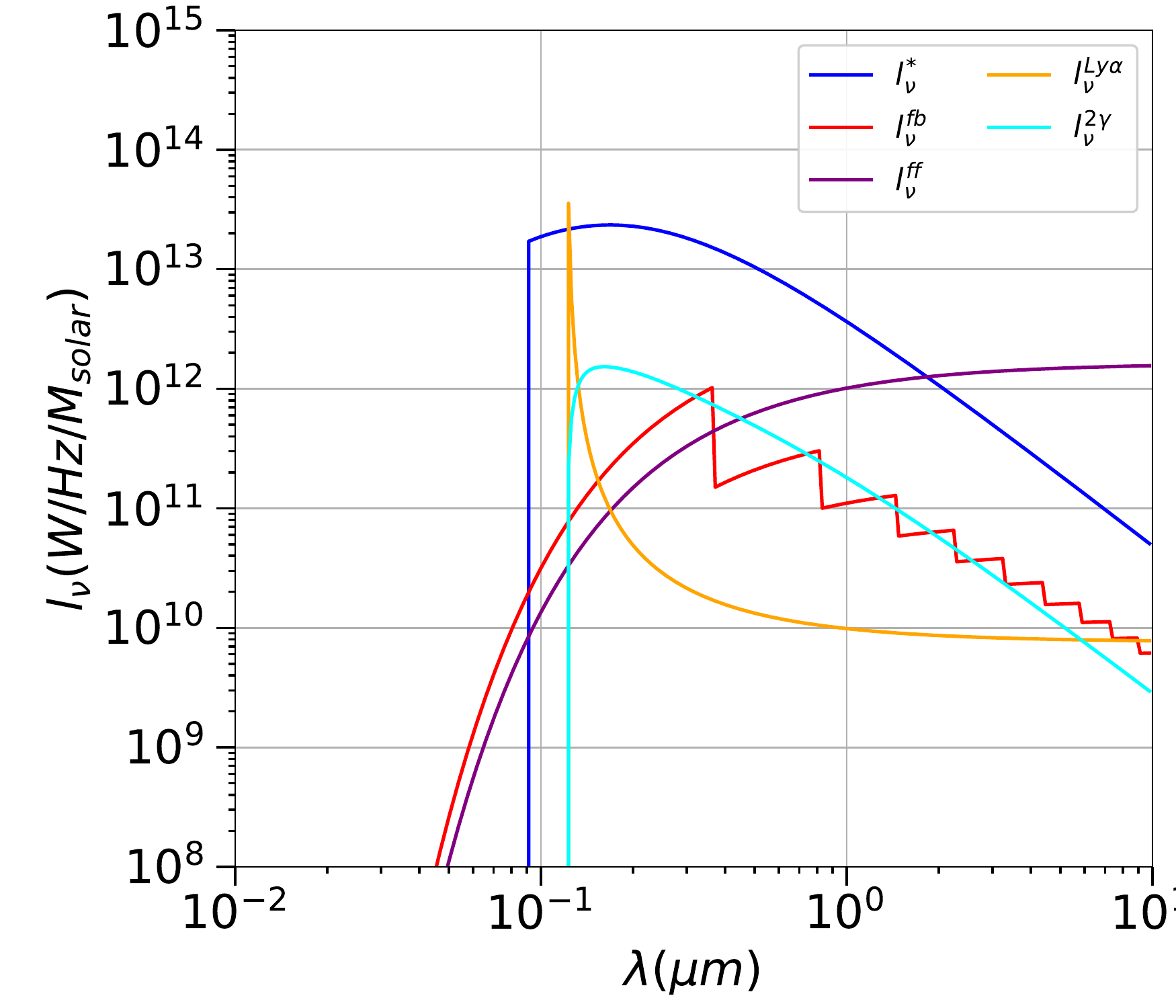}\caption{The different contributions to the luminosity mass density $l_{\nu}$ as a function of the rest-frame wavelength $\lambda$ for stars at $z= 10$. The most important contribution (blue line) comes from the stellar spectrum. Red and purple lines indicate the free-free and free-bounds contributions, while the yellow line is the Lyman-$\alpha$ luminosity. Finally, the cyan line is the contribution from two-photon processes.} 
\label{fig:lnu}
\end{figure}

For each emission process described above we derive the mean luminosity by integrating over the stellar initial mass function (IMF) $f(M_*)$:
\begin{equation}
    l_{\nu} = \frac{\int dM_* f(M_*) L_{\nu}(M_*)}{\int dM_* M_* f(M_*)}
\end{equation}
As mentioned above, we focus on metal-poor stars since they provide a larger contribution to the IR-background than metal-free stars, and we adopt the IMF given by Salpeter~\cite{SalpeterMass} $f(M_*) \propto M_*^{-2.35}$ in the mass range between 3 and 150 $M_{\odot}$. As an example, in Fig.~\ref{fig:lnu} we show the different contributions to the mean luminosity from Pop II stars at $z = 10$.

The weight function for high-z galaxies is obtained summing the various contributions described in this section and using Eq.~\ref{eq:Whighz}, where the mean stellar lifetime $\langle\tau_* \rangle$ for metal-poor stars is $\langle \tau_{*} \rangle = \int dM_* f(M_*) \tau_*(M_*)$,
with $\log_{10}(\tau_{*}(M_*)/\text{yr}) =9.59 -2.79 x + 0.63 x^2.$

\section{Best-fit parameters} \label{app:fit}

In Tab. \ref{tab:param_HSTonly} and Tab. \ref{tab:paramSpitzerOnly} we report the best fit parameters for all analyzed datasets. In the main text, for brevity, we reported in Tab.\ref{tab:param_HSTSpitzer} only the results of the combined datasets HST-Spitzer. Here we also show the complete results for the HST-only and Spitzer-only cases. 

\begin{table}
  \centering
  \renewcommand{\arraystretch}{2}
\begin{tabular}[7pt]{c | c | c | c}
        \hline\hline
        Parameters & ALPs + IHL  & ALPs-only & Prior ranges\\
        \hline
        $m_a$ & $2.50^{+0.22}_{-0.16}$ eV &  $2.58^{+0.20}_{-0.18}$ eV & $(0, 10) $ eV\\
        $\tau_a$ & $(5.03^{+0.45}_{-0.42})\times 10^{23}$ s & $(4.08^{+0.26}_{-0.30})\times 10^{23}$ s & $(10^{20}, 10^{28})$ s \\
        $A_{IHL}$ & $ 0.969^{+0.129}_{-0.126}$ & -- & $(0, 10^3)$ \\
        $A_{06}$ & $(1.32^{+0.17}_{-0.17}) \times 10^3$ & $(1.91^{+0.16}_{-0.15}) \times 10^3 $ & $(0, 10^7)$\\
        $A_{07}$ & $ (2.53^{+0.32}_{-0.32}) \times 10^3 $ &  $ (3.60^{+0.30}_{-0.29})\times 10^3 $ & $(0, 10^7)$ \\ $A_{08}$ & $ (2.31^{+0.32}_{-0.32})\times 10^3 $ & $( 3.86^{+0.24}_{-0.25}) \times 10^3$ & $(0, 10^7)$ \\ $A_{12}$ & $( 9.48^{+6.16}_{-5.32}) \times 10^3$ & $ (3.83^{+6.05}_{-3.82}) \times 10^3 $ & $(0, 10^7)$ \\ $A_{16} $ & $ (1.68^{+0.46}_{-0.51}) \times 10^4$ & $(1.68^{+0.46}_{-0.52}) \times 10^4 $ & $(0, 10^7)$  \\ $\chi^2(\chi^2_{\text{reduced}})$ & $116.2 (1.9) $ & $59.2 (0.97)$ & -- \\
        \hline\hline
    \end{tabular}
    \caption{\textbf{Best fit HST only.} Best fit values with $1\sigma$ intervals for the HST only analysis, with or without the IHL contribution.} \label{tab:param_HSTonly}
\end{table}

\begin{table}
  \centering
  \renewcommand{\arraystretch}{2}
\begin{tabular}[7pt]{c | c | c | c}
        \hline\hline
        Parameters & ALPs + IHL  & ALPs-only & Prior ranges\\
        \hline
        $m_a$ & $0.80^{+0.03}_{-0.03}$ eV &  $2.83^{+0.03}_{-0.03}$ eV & $(0, 10) $ eV\\
        $\tau_a$ & $(8.58^{+0.79}_{-0.77})\times 10^{24}$ s & $(7.80^{+0.56}_{-0.66})\times 10^{24}$ s & $(10^{20}, 10^{28})$ s \\
        $A_{IHL}$ & $ 0.020^{+0.029}_{-0.015}$ & -- & $(0, 10^3)$ \\
        $A_{36}$ & $ 3.20^{+3.31}_{-2.26}$ & $ 3.18^{+3.31}_{-2.16} $ & $(0, 10^3)$ \\ $A_{45} $ & $ 18.8^{+3.3}_{-3.2}$ & $19.5^{+3.5}_{-3.2}$ & $(0, 10^3)$ \\ $\chi^2(\chi^2_{\text{reduced}})$ & $26.2 (0.6)$ & $26.2 (0.6)$ & -- \\  
        \hline\hline
    \end{tabular}
    \caption{\textbf{Best fit Spitzer only.} Best fit values with $1\sigma$ intervals for the Spitzer only analysis, with or without the IHL contribution.} \label{tab:paramSpitzerOnly}
\end{table}

\newpage
\bibliographystyle{hunsrt}
\bibliography{biblio}

\begin{thebibliography}{10}

\bibitem{Peccei:1977hh}
R.D. Peccei and Helen~R. Quinn.
\newblock {CP Conservation in the Presence of Instantons}.
\newblock {\em Phys. Rev. Lett.}, 38:1440--1443, 1977.

\bibitem{Peccei:1977ur}
R.~D. Peccei and Helen~R. Quinn.
\newblock {Constraints Imposed by CP Conservation in the Presence of
  Instantons}.
\newblock {\em Phys. Rev.}, D16:1791--1797, 1977.

\bibitem{Weinberg:1977ma}
Steven Weinberg.
\newblock {A New Light Boson?}
\newblock {\em Phys. Rev. Lett.}, 40:223--226, 1978.

\bibitem{Wilczek:1977pj}
Frank Wilczek.
\newblock {Problem of Strong p and t Invariance in the Presence of Instantons}.
\newblock {\em Phys. Rev. Lett.}, 40:279--282, 1978.

\bibitem{Abbott:1982af}
L.F. Abbott and P.~Sikivie.
\newblock {A Cosmological Bound on the Invisible Axion}.
\newblock {\em Phys. Lett. B}, 120:133--136, 1983.

\bibitem{Preskill:1982cy}
John Preskill, Mark~B. Wise, and Frank Wilczek.
\newblock {Cosmology of the Invisible Axion}.
\newblock {\em Phys. Lett. B}, 120:127--132, 1983.

\bibitem{Dine:1981rt}
Michael Dine, Willy Fischler, and Mark Srednicki.
\newblock {A Simple Solution to the Strong CP Problem with a Harmless Axion}.
\newblock {\em Phys. Lett.}, 104B:199--202, 1981.

\bibitem{Dine:1982ah}
Michael Dine and Willy Fischler.
\newblock {The Not So Harmless Axion}.
\newblock {\em Phys. Lett. B}, 120:137--141, 1983.

\bibitem{DiLuzio:2016sbl}
Luca Di~Luzio, Federico Mescia, and Enrico Nardi.
\newblock {Redefining the Axion Window}.
\newblock {\em Phys. Rev. Lett.}, 118(3):031801, 2017, 1610.07593.

\bibitem{Caputo:2018vmy}
Andrea Caputo, Marco Regis, Marco Taoso, and Samuel~J. Witte.
\newblock {Detecting the Stimulated Decay of Axions at Radio Frequencies}.
\newblock {\em JCAP}, 03:027, 2019, 1811.08436.

\bibitem{Caputo:2018ljp}
Andrea Caputo, Carlos~Peña Garay, and Samuel~J. Witte.
\newblock {Looking for axion dark matter in dwarf spheroidal galaxies}.
\newblock {\em Phys. Rev.}, D98(8):083024, 2018, 1805.08780.

\bibitem{Kephart:1994uy}
Thomas~W. Kephart and Thomas~J. Weiler.
\newblock {Stimulated radiation from axion cluster evolution}.
\newblock {\em Phys. Rev.}, D52:3226--3238, 1995.

\bibitem{Rosa:2017ury}
João~G. Rosa and Thomas~W. Kephart.
\newblock {Stimulated Axion Decay in Superradiant Clouds around Primordial
  Black Holes}.
\newblock {\em Phys. Rev. Lett.}, 120(23):231102, 2018, 1709.06581.

\bibitem{Tkachev:1987cd}
I.~I. Tkachev.
\newblock {An Axionic Laser in the Center of a Galaxy?}
\newblock {\em Phys. Lett.}, B191:41--45, 1987.

\bibitem{Sigl:2019pmj}
G\"unter Sigl and Pranjal Trivedi.
\newblock {Axion Condensate Dark Matter Constraints from Resonant Enhancement
  of Background Radiation}.
\newblock 7 2019, 1907.04849.

\bibitem{Battye:2019aco}
Richard~A. Battye, Bjoern Garbrecht, Jamie~I. McDonald, Francesco Pace, and
  Sankarshana Srinivasan.
\newblock {Dark matter axion detection in the radio/mm-waveband}.
\newblock {\em Phys. Rev. D}, 102(2):023504, 2020, 1910.11907.

\bibitem{Arza:2019nta}
Ariel Arza and Pierre Sikivie.
\newblock {Production and detection of an axion dark matter echo}.
\newblock {\em Phys. Rev. Lett.}, 123(13):131804, 2019, 1902.00114.

\bibitem{Ghosh:2020hgd}
Oindrila Ghosh, Jordi Salvado, and Jordi Miralda-Escud\'e.
\newblock {Axion Gegenschein: Probing Back-scattering of Astrophysical Radio
  Sources Induced by Dark Matter}.
\newblock 8 2020, 2008.02729.

\bibitem{Creque-Sarbinowski:2018ebl}
Cyril Creque-Sarbinowski and Marc Kamionkowski.
\newblock {Searching for Decaying and Annihilating Dark Matter with Line
  Intensity Mapping}.
\newblock {\em Phys. Rev.}, D98(6):063524, 2018, 1806.11119.

\bibitem{Arias:2012az}
Paola Arias, Davide Cadamuro, Mark Goodsell, Joerg Jaeckel, Javier Redondo, and
  Andreas Ringwald.
\newblock {WISPy Cold Dark Matter}.
\newblock {\em JCAP}, 06:013, 2012, 1201.5902.

\bibitem{Grin:2006aw}
Daniel Grin, Giovanni Covone, Jean-Paul Kneib, Marc Kamionkowski, Andrew Blain,
  and Eric Jullo.
\newblock {A Telescope Search for Decaying Relic Axions}.
\newblock {\em Phys. Rev.}, D75:105018, 2007, astro-ph/0611502.

\bibitem{Ressell:1991zv}
M.~Ted Ressell.
\newblock {Limits to the radiative decay of the axion}.
\newblock {\em Phys. Rev.}, D44:3001--3020, 1991.

\bibitem{Bershady:1990sw}
Matthew~A. Bershady, M.~Ted Ressell, and Michael~S. Turner.
\newblock {Telescope search for multi-eV axions}.
\newblock {\em Phys. Rev. Lett.}, 66:1398--1401, 1991.

\bibitem{Bernal:2020lkd}
Jos\'e~Luis Bernal, Andrea Caputo, and Marc Kamionkowski.
\newblock {Strategies to Detect Dark-Matter Decays with Line-Intensity
  Mapping}.
\newblock 12 2020, 2012.00771.

\bibitem{Dwek:2005dj}
Eli Dwek, Richard~G. Arendt, and Frank Krennrich.
\newblock {The Near infrared background: Interplanetary dust or primordial
  stars?}
\newblock {\em Astrophys. J.}, 635:784--794, 2005, astro-ph/0508262.

\bibitem{Kohri:2017ljt}
Kazunori Kohri and Hideo Kodama.
\newblock {Axion-Like Particles and Recent Observations of the Cosmic Infrared
  Background Radiation}.
\newblock {\em Phys. Rev. D}, 96(5):051701, 2017, 1704.05189.

\bibitem{Kohri:2017oqn}
Kazunori Kohri, Takeo Moroi, and Kazunori Nakayama.
\newblock {Can decaying particle explain cosmic infrared background excess?}
\newblock {\em Phys. Lett.}, B772:628--633, 2017, 1706.04921.

\bibitem{Kalashev:2018bra}
Oleg~E. Kalashev, Alexander Kusenko, and Edoardo Vitagliano.
\newblock {Cosmic infrared background excess from axionlike particles and
  implications for multimessenger observations of blazars}.
\newblock {\em Phys. Rev. D}, 99(2):023002, 2019, 1808.05613.

\bibitem{2018RvMP...90b5006K}
A.~{Kashlinsky}, R.~G. {Arendt}, F.~{Atrio-Barandela}, N.~{Cappelluti},
  A.~{Ferrara}, and G.~{Hasinger}.
\newblock {Looking at cosmic near-infrared background radiation anisotropies}.
\newblock {\em Reviews of Modern Physics}, 90(2):025006, April 2018,
  1802.07774.

\bibitem{Gong:2015hke}
Yan Gong, Asantha Cooray, Ketron Mitchell-Wynne, Xuelei Chen, Michael Zemcov,
  and Joseph Smidt.
\newblock {Axion decay and anisotropy of near-IR extragalactic background
  light}.
\newblock {\em Astrophys. J.}, 825(2):104, 2016, 1511.01577.

\bibitem{Regis:2020fhw}
Marco Regis, Marco Taoso, Daniel Vaz, Jarle Brinchmann, Sebastiaan~L.
  Zoutendijk, Nicolas Bouch\'e, and Matthias Steinmetz.
\newblock {Searching for Light in the Darkness: Bounds on ALP Dark Matter with
  the optical MUSE-Faint survey}.
\newblock 9 2020, 2009.01310.

\bibitem{1953ApJ...117..134L}
D.~N. {Limber}.
\newblock {The Analysis of Counts of the Extragalactic Nebulae in Terms of a
  Fluctuating Density Field.}
\newblock {\em ApJ}, 117:134, January 1953.

\bibitem{Mitchell_Wynne_2015}
Ketron Mitchell-Wynne, Asantha Cooray, Yan Gong, Matthew Ashby, Timothy Dolch,
  Henry Ferguson, Steven Finkelstein, Norman Grogin, Dale Kocevski, Anton
  Koekemoer, and et~al.
\newblock Ultraviolet luminosity density of the universe during the epoch of
  reionization.
\newblock {\em Nature Communications}, 6(1), Sep 2015.

\bibitem{Windhorst:2010ib}
Rogier~A. Windhorst et~al.
\newblock {The Hubble Space Telescope Wide Field Camera 3 Early Release Science
  data: Panchromatic Faint Object Counts for 0.2-2 microns wavelength}.
\newblock {\em Astrophys. J. Suppl.}, 193:27, 2011, 1005.2776.

\bibitem{Helgason_2012}
Kari Helgason, Massimo Ricotti, and Alexander Kashlinsky.
\newblock Reconstructing the near-infrared background fluctuations from known
  galaxy populations using multiband measurements of luminosity functions.
\newblock {\em The Astrophysical Journal}, 752(2):113, Jun 2012.

\bibitem{KASHLINSKY_2005}
A~Kashlinsky.
\newblock Cosmic infrared background and early galaxy evolution.
\newblock {\em Physics Reports}, 409(6):361–438, Apr 2005.

\bibitem{Kashlinsky_2007}
Alexander Kashlinsky, R.G. Arendt, John~C. Mather, and S.H. Moseley.
\newblock {New measurements of cosmic infrared background fluctuations from
  early epochs}.
\newblock {\em Astrophys. J. Lett.}, 654:L5--L8, 2007, astro-ph/0612445.
\newblock [Erratum: Astrophys.J.Lett. 657, L131 (2007), Erratum: Astrophys.J.
  657, L131 (2007)].

\bibitem{Cooray_2012}
Asantha Cooray, Yan Gong, Joseph Smidt, and Mario~G. Santos.
\newblock The near-infrared background intensity and anisotropies during the
  epoch of reionization.
\newblock {\em The Astrophysical Journal}, 756(1):92, Aug 2012.

\bibitem{Santos_2002}
M.~R. Santos, V.~Bromm, and M.~Kamionkowski.
\newblock The contribution of the first stars to the cosmic infrared
  background.
\newblock {\em Monthly Notices of the Royal Astronomical Society},
  336(4):1082–1092, Nov 2002.

\bibitem{Schneider:2018xba}
Aurel Schneider.
\newblock {Constraining noncold dark matter models with the global 21-cm
  signal}.
\newblock {\em Phys. Rev. D}, 98(6):063021, 2018, 1805.00021.

\bibitem{Wise:2014vwa}
John~H. Wise, Vasiliy~G. Demchenko, Martin~T. Halicek, Michael~L. Norman,
  Matthew~J. Turk, Tom Abel, and Britton~D. Smith.
\newblock {The birth of a galaxy \textendash{} III. Propelling reionization
  with the faintest galaxies}.
\newblock {\em Mon. Not. Roy. Astron. Soc.}, 442(3):2560--2579, 2014,
  1403.6123.

\bibitem{Sheth:1999mn}
Ravi~K. Sheth and Giuseppe Tormen.
\newblock {Large scale bias and the peak background split}.
\newblock {\em Mon.Not.Roy.Astron.Soc.}, 308:119, 1999, astro-ph/9901122.

\bibitem{Diemer_2018}
Benedikt Diemer.
\newblock Colossus: A python toolkit for cosmology, large-scale structure, and
  dark matter halos.
\newblock {\em The Astrophysical Journal Supplement Series}, 239(2):35, Dec
  2018.

\bibitem{cooray2012measurement}
Asantha Cooray, Joseph Smidt, Francesco~De Bernardis, Yan Gong, Daniel Stern,
  Matthew L.~N. Ashby, Peter~R. Eisenhardt, Christopher~C. Frazer, Anthony~H.
  Gonzalez, Christopher~S. Kochanek, Szymon Kozlowski, and Edward~L. Wright.
\newblock A measurement of the intrahalo light fraction with near-infrared
  background anisotropies, 2012, 1210.6031.

\bibitem{Lin:2004ak}
Yen-Ting Lin, Joseph~J. Mohr, and S.~Adam Stanford.
\newblock {K-band properties of galaxy clusters and groups: Luminosity
  function, radial distribution and halo occupation number}.
\newblock {\em Astrophys. J.}, 610:745--761, 2004, astro-ph/0402308.

\bibitem{Fornengo:2013rga}
Nicolao Fornengo and Marco Regis.
\newblock {Particle dark matter searches in the anisotropic sky}.
\newblock {\em Front. Physics}, 2:6, 2014, 1312.4835.

\bibitem{Zheng:2004id}
Zheng Zheng, Andreas~A. Berlind, David~H. Weinberg, Andrew~J. Benson,
  Carlton~M. Baugh, et~al.
\newblock {Theoretical models of the halo occupation distribution: Separating
  central and satellite galaxies}.
\newblock {\em Astrophys.J.}, 633:791--809, 2005, astro-ph/0408564.

\bibitem{Kashlinsky:2005di}
Alexander Kashlinsky, R.G. Arendt, John~C. Mather, and S.H. Moseley.
\newblock {Tracing the first stars with fluctuations of the cosmic infrared
  background}.
\newblock {\em Nature}, 438:45--50, 2005, astro-ph/0511105.

\bibitem{Kashlinsky:2012zz}
A.~Kashlinsky, R.G. Arendt, M.L.N. Ashby, G.G. Fazio, J.~Mather, and S.H.
  Moseley.
\newblock {New measurements of the cosmic infrared background fluctuations in
  deep Spitzer/IRAC survey data and their cosmological implications}.
\newblock {\em Astrophys. J.}, 753:63, 2012, 1201.5617.

\bibitem{Seo:2015fga}
H.J. Seo, Hyung~Mok Lee, T.~Matsumoto, W.~S. Jeong, Myung~Gyoon Lee, and
  J.~Pyo.
\newblock {AKARI Observation of the Sub-degree Scale Fluctuation of the
  Near-infrared Background}.
\newblock {\em Astrophys. J.}, 807(2):140, 2015, 1504.05681.

\bibitem{Bock:2012fw}
J.~Bock et~al.
\newblock {The Cosmic Infrared Background Experiment (CIBER): The Wide-Field
  Imagers}.
\newblock {\em Astrophys. J. Suppl.}, 207:32, 2013, 1206.4702.

\bibitem{Helgason:2016xoc}
K\'ari Helgason and Eiichiro Komatsu.
\newblock {AKARI near-infrared background fluctuations arise from normal galaxy
  populations}.
\newblock {\em Mon. Not. Roy. Astron. Soc.}, 467(1):L36--L40, 2017, 1611.00042.

\bibitem{10.1093/biomet/57.1.97}
W.~K. Hastings.
\newblock {Monte Carlo sampling methods using Markov chains and their
  applications}.
\newblock {\em Biometrika}, 57(1):97--109, 04 1970,
  https://academic.oup.com/biomet/article-pdf/57/1/97/23940249/57-1-97.pdf.

\bibitem{Metropolis1953}
Nicholas Metropolis, Arianna~W. Rosenbluth, Marshall~N. Rosenbluth, Augusta~H.
  Teller, and Edward Teller.
\newblock Equation of state calculations by fast computing machines.
\newblock {\em The Journal of Chemical Physics}, 21(6):1087--1092, 1953.

\bibitem{Foreman_Mackey_2013}
Daniel Foreman-Mackey, David~W. Hogg, Dustin Lang, and Jonathan Goodman.
\newblock emcee: The mcmc hammer.
\newblock {\em Publications of the Astronomical Society of the Pacific},
  125(925):306–312, Mar 2013.

\bibitem{Ayala:2014pea}
Adrian Ayala, Inma Dom\'\i{}nguez, Maurizio Giannotti, Alessandro Mirizzi, and
  Oscar Straniero.
\newblock {Revisiting the bound on axion-photon coupling from Globular
  Clusters}.
\newblock {\em Phys. Rev. Lett.}, 113(19):191302, 2014, 1406.6053.

\bibitem{Bloch:2020uzh}
Itay~M. Bloch, Andrea Caputo, Rouven Essig, Diego Redigolo, Mukul Sholapurkar,
  and Tomer Volansky.
\newblock {Exploring New Physics with O(keV) Electron Recoils in Direct
  Detection Experiments}.
\newblock 6 2020, 2006.14521.

\bibitem{Masso:2005ym}
Eduard Masso and Javier Redondo.
\newblock {Evading astrophysical constraints on axion-like particles}.
\newblock {\em JCAP}, 09:015, 2005, hep-ph/0504202.

\bibitem{Jaeckel_2007}
Joerg Jaeckel, Eduard Massó, Javier Redondo, Andreas Ringwald, and Fuminobu
  Takahashi.
\newblock Need for purely laboratory-based axionlike particle searches.
\newblock {\em Physical Review D}, 75(1), Jan 2007.

\bibitem{Li_2018}
Yanxia Li, Nico Cappelluti, Richard~G. Arendt, Günther Hasinger, Alexander
  Kashlinsky, and Kari Helgason.
\newblock The splash and chandra cosmos legacy survey: The cross-power between
  near-infrared and x-ray background fluctuations.
\newblock {\em The Astrophysical Journal}, 864(2):141, Sep 2018.

\bibitem{Mitchell-Wynne:2016hgd}
Ketron Mitchell-Wynne, Asantha Cooray, Yongquan Xue, Bin Luo, William Brandt,
  and Anton Koekemoer.
\newblock {Cross-correlation between X-ray and optical/near-infrared background
  intensity fluctuations}.
\newblock {\em Astrophys. J.}, 832(2):104, 2016, 1610.02015.

\bibitem{Cappelluti:2017nww}
N.~Cappelluti, R.~Arendt, A.~Kashlinsky, Y.~Li, G.~Hasinger, K.~Helgason,
  M.~Urry, P.~Natarajan, and A.~Finoguenov.
\newblock {Probing large scale coherence between Spitzer IR and Chandra X-ray
  source-subtracted cosmic backgrounds}.
\newblock {\em Astrophys. J. Lett.}, 847(1):L11, 2017, 1709.02824.

\bibitem{Cappelluti_2013}
N.~Cappelluti, A.~Kashlinsky, R.~G. Arendt, A.~Comastri, G.~G. Fazio,
  A.~Finoguenov, G.~Hasinger, J.~C. Mather, T.~Miyaji, and S.~H. Moseley.
\newblock Cross-correlating cosmic infrared and x-ray background fluctuations:
  Evidence of significant black hole populations among the cib sources.
\newblock {\em The Astrophysical Journal}, 769(1):68, May 2013.

\bibitem{Fernandez2006}
Elizabeth~R. {Fernandez} and Eiichiro {Komatsu}.
\newblock {The Cosmic Near-Infrared Background: Remnant Light from Early
  Stars}.
\newblock {\em The Astrophysical Journal}, 646(2):703--718, August 2006,
  astro-ph/0508174.

\bibitem{Canta}
Sebastiano {Cantalupo}, Cristiano {Porciani}, and Simon~J. {Lilly}.
\newblock {Mapping Neutral Hydrogen during Reionization with the
  Ly{\ensuremath{\alpha}} Emission from Quasar Ionization Fronts}.
\newblock {\em The Astrophysical Journal}, 672(1):48--58, January 2008,
  0709.0654.

\bibitem{Santos2002}
Michael~R. {Santos}, Volker {Bromm}, and Marc {Kamionkowski}.
\newblock {The contribution of the first stars to the cosmic infrared
  background}.
\newblock 336(4):1082--1092, November 2002, astro-ph/0111467.

\bibitem{SalpeterMass}
Edwin~E. {Salpeter}.
\newblock {The Luminosity Function and Stellar Evolution.}
\newblock {\em The Astrophysical Journal}, 121:161, January 1955.

\end{thebibliography}

\end{document}